\documentclass[reprint,amsmath,amssymb,aps]{revtex4-2}

\usepackage{amsmath,amsfonts,amssymb}
\usepackage{upgreek}
\usepackage{float}
\usepackage[colorlinks=true,linkcolor=blue,citecolor=blue,urlcolor=blue]{hyperref}
\usepackage{graphicx}
\usepackage{dcolumn}
\usepackage{bm}
\usepackage{xfrac}
\begin{document}

\title{High-order synchronization in identical  neurons with asymmetric pulse coupling}

\author{Abhay}
\email{p20180031@goa.bits-pilani.ac.in,abhayphysicsphd@gmail.com}
\author{Gaurav Dar}
\email{gdar@goa.bits-pilani.ac.in}
\affiliation{Department of Physics, BITS Pilani K K Birla Goa Campus, Zuarinagar, Goa-403726, India}

\date{\today}

\begin{abstract}
The phenomenon of high-order ($p/q$) synchronization, induced by \textit{two different} frequencies in the system, is well-known and studied extensively in forced oscillators including neurons and to a lesser extent in coupled oscillators. Their frequencies are locked such that for every $p$ cycles of one oscillator there are $q$ cycles of the other. We demonstrate this phenomenon in a pair of coupled neurons having \textit{identical} frequencies but \textit{asymmetric} coupling.
Specifically, we focus on an excitatory(E)-inhibitory(I) neuron pair where such an asymmetry is naturally present even with equal reciprocal synaptic strengths $(g)$ and inverse time constant $(\alpha)$.
We thoroughly investigate the asymmetric coupling-induced $p/q$ frequency locking structure in $(g,\alpha)$ parameter space through simulations and analysis. Simulations display quasiperiodicity, devil staircase, a novel Farey arrangement of spike sequences, and presence of reducible and irreducible $p/q$ regions. We introduce an analytical method, based on event-driven maps, 
to determine the existence and stability of any spike sequence of the two neurons in  a $p/q$ frequency-locked state. Specifically, this method successfully deals with non-smooth bifurcations and we could utilize it to obtain solutions for the case of identical E-I neuron pair under arbitrary coupling strength. In contrast to the so-called Arnold tongues, the
$p/q$ regions obtained here are not structure-less. Instead they have their own internal bifurcation structure with varying levels of complexity. Intra-sequence and inter-sequence multistability, involving spike sequences of same $p/q$ state, are found. Additionally, multistability also arises by overlap of $p/q$ with $p'/q'$.
The boundaries of both reducible and irreducible $p/q$ regions are defined by saddle node and non-smooth grazing bifurcations of various types.
\end{abstract}

\maketitle

\section{Introduction}
\label{sec:Intro}
The phenomenon of synchronization pertaining to the alignment of frequencies is commonly observed in systems of forced and interconnected oscillators.  This alignment ranges from simple frequency matching, also known as $1/1$ synchronization or frequency-locked states, to more intricate relationships characterized by rational frequency ratios, known as high-order synchronization ($p/q$ frequency-locked states).
Simple and high-order frequency locking is pervasive in forced oscillators examples of which include \textit{forced} nano-oscillators \cite{Shim2007},  chemical oscillators\cite{Vance1989,Aronson1986,Nakata2009}, 
Lasers\cite{Simonet1994,Mendez2001}, electronic oscillators\cite{VANDERPOL1927,METTIN1993,Behta2023,Flaherty1978}, blinking fireflies\cite{Buck1981}, heart cells \cite{Guevara1988,Glass1983}, auditory hair cells \cite{FredricksonHemsing2012,Levy2016}, corticothalamic system\cite{Alinejad2020}, and Josephson junctions\cite{Garcalvarez2008}.
Simple $1/1$ frequency locking in autonomous \textit{mutually coupled} oscillators is also extensively documented in mechanical systems\cite {Pantaleone2002,Oliveira2015,Lu2023,Srikanth2022}, nano oscillators\cite{Colombano2019,Matheny2014,Singh2019nano,Kaka2005}, chemical reactions\cite{Nakata1998,Neu1979,Wickramasinghe2013,Marek1975,Tajima2016,Cohen1979,Liu2022}, electronic circuits\cite{Ignatov2016,English2015,Parihar2015,PaulAsir2022,Dixit2019,Singh2019,Ivanchenko2004}, Lasers\cite{Wnsche2005}, and biological systems\cite{GholizadeNarm2009,Bier2000}. However, instances of high-order synchronization are comparatively uncommon in existing literature. Nevertheless, occurrences of it have been observed in an experiment involving coupled electrochemical oscillators \cite{Montoya2013}, an electronic circuit simulating coupled pacemakers \cite{Ypey1980}, a theoretical study involving coupled Van der Pol oscillators \cite{Schilder2007}, Josephson junctions\cite{Valkering2000},  and a few set of studies involving mutually coupled neurons \cite{Coombes1999,Bressloff1998,Ermentrout1981}.
\par
Studying synchronization in neurons is crucial in neuroscience as it underlies fundamental brain functions such as perception, cognition, and motor coordination \cite{Ward2003,Buzsaki2004}. While considerable attention has been devoted to simple $1/1$ frequency locking across various \textit{coupled} neuron models such as Hodgkin-Huxley \cite{Hansel1993,Rossoni2005}, Morris-Leccar\cite{ermentrout2010mathematical}, Connor-Steven \cite{Hansel1995}, reduced neuron models\cite{Buri2003,YongliSong2012,FarajzadehTehrani2015}, and integrate-and-fire models \cite{van_vreeswijk_1994,van1996partial,Chartrand2019,Ermentrout1996,Pfeuty2003,Ratas2022}, high-order synchronization has been largely neglected in literature, with only a sparse number of studies addressing it. Notably, initial observations were made in an experimental setup involving two pacemaker cells coupled through electrical synapses \cite{Jalife1984} and confirmed in simulations using a conductance-based neuron model \cite{Ypey1980}. Furthermore,  coupled leaky integrate-and-fire (LIF) neurons are shown to exhibit a few high-order $p/q$ frequency locking under pulse coupling through chemical synapses \cite{Coombes1999,Bressloff1998,Chow1998}.
\par
The various high-order frequency-locked regions tend to be organized in \lq Arnold tongues' in the space of frequency detuning and coupling strength \cite{pikovsky2001synchronization}. A $p/q$ tongue originates from points of zero coupling strength and frequency ratio equal to $p/q$ expanding as the coupling strength increases. Notably, a frequency mismatch close to zero yields $1/1$ locking, while $p/q$ locking occurs for large mismatch. This behavior is effectively modeled by the circle map\cite{pikovsky2001synchronization}. Distinct from the above scenario was the finding of $1/2$ in a pair of inhibitory neurons with identical frequencies \cite{Bressloff1998}. In another recent study, a pair of Stuart-Landau oscillators coupled through asymmetric quadratic coupling displayed a consistent $1/2$ frequency locking regardless of frequency mismatch \cite{Thomas2021}. However, the extensive range and complexity of high-order frequency locking observed through the Arnold tongues has not been demonstrated in any study of \textit{identical} coupled oscillators. We show that coupled neurons with \textit{identical} frequencies under an \textit{asymmetric} coupling can display a rich tapestry of  high-order frequency lockings.
\par
Several studies investigate the effect of asymmetric coupling on various aspects of the dynamics of interconnected oscillators. For instance, coupling asymmetry can regulate multistability in Van der Pol oscillators \cite{Astakhov2016}, lead to the coexistence of periodic and chaotic attractors in Hindmarsh-Rose neurons \cite{Pisarchik2018,TabekouengNjitacke2020}, and modifies bifurcations involving in-phase and out-of-phase 1/1 states in pulse-coupled integrate-and-fire neurons \cite{Zeitler2009}. Additional aspects of such asymmetry have been investigated in \cite{Blasius2005,Bragard2007,Pinto2006,Ryu2020,Astakhov2013}. However, potential occurrence of \textit{high-order} $p/q$ frequency locking among \textit{identical} oscillators through \textit{asymmetric} coupling has been overlooked. In an excitatory(E)-inhibitory(I) neuron pair asymmetry in coupling by chemical synapses occurs naturally.  High-order synchronization in such an E-I pair of LIF neurons has been explored for \textit{non-identical} neurons \cite{Bressloff1998}. Our aim is to explore the role of \textit{coupling asymmetry} alone in inducing high-order synchronization amongst \textit{identical} neurons.
\par
On the boundaries of Arnold tongues, various types of smooth bifurcations can occur, including saddle-node bifurcations\cite{pikovsky2001synchronization}, period-doubling bifurcations\cite{hilborn2000chaos}, Hopf-bifurcation\cite{FredricksonHemsing2012} and Neimark-Sacker bifurcations\cite{METTIN1993}. These bifurcations play a crucial role in shaping the dynamics of coupled oscillators and can lead to emergence of complex behaviors such as multistability between distinct  $p/q$ and $p'/q'$ frequency-locked states.
However, existence of bifurcations \textit{inside} a $p/q$ tongue involving only its own states, are unknown. Our study on coupled neuronal pairs illustrate a unusual result: we observe bifurcation structures \textit{inside} a tongue involving states of the same $p/q$. 
\par
Integrate-and-fire (IF) oscillators are threshold systems with non-smooth dynamics. The discontinuity in the dynamics of non-smooth systems are known to induce several types of non-smooth bifurcations \cite{Banerjee1999,Nusse1992,JAIN2003,Casas1996,Makarenkov2012}. Analytical studies of several periodically \textit{forced} IF neuron models have revealed that the boundaries of resulting Arnold tongues are shaped by a combination of smooth and non-smooth bifurcations \cite{KhajehAlijani2009,Coombes2012,Coombes2001,LAING2005}. In these analytical formulations, which utilize firing time maps, additional conditions are incorporated to address the loss of $p/q$ frequency-locked solutions due to non-smooth bifurcations.
However, existing analytical formulations for \textit{coupled} IF neurons \cite{Coombes1999,van_vreeswijk_1994,Chow1998} lack such conditions. Without these considerations, there is a potential for generating unphysical frequency-locked solutions and overlooking non-smooth bifurcations. This study will address this gap by developing an alternate formalism for the analysis of coupled LIF neurons.
\par
We employ event-driven maps of LIF neurons to devise an analytical approach for exploring frequency locking in a pair of pulse-coupled LIF neurons. These maps establish a connection between the states of neurons across successive network spikes, achieved by integrating the model equations between consecutive spikes while replacing the summation over infinite spikes with a differential equation for synaptic current. Event-driven maps have been previously utilized to accelerate simulations of large neuronal networks, including sparse \cite{Luccioli2012}, random\cite{Tattini2012}, and globally coupled networks of excitatory neurons \cite{Zillmer2007}. Splay states could be analytically obtained from these maps, in a straightforward manner, for a globally coupled network of neurons \cite{Zillmer2007,Olmi2012,Zillmer2006}. Additionally, a linear stability analysis for splay states was derived in the same reference from event-driven maps and examined under thermodynamic limit. We construct a systematic procedure using event driven maps to find the regions of existence and stability of $p/q$ frequency - locked states in parameter space of synaptic strength and inverse time constant. Crucially, our procedure incorporates conditions that lead to non-smooth bifurcations and we use these to obtain results for an E-I pair of \textit{identical} LIF neurons with no constraints on synaptic strength.
\par
 The paper is organized as follows: In Section \ref{sec:Eqn_of_motion}, we describe event-driven maps for a pair of pulse coupled LIF neurons, specifically focusing on an excitatory-inhibitory neuron pair. We further describe the physical quantities of our interest. In Section \ref{sec:simulation_results}, we describe simulation results for high-order frequency locking in an E-I pair of identical neurons. In Section \ref{sec:pbyq_freq_locked_analytical}, we develop an analytical method to study the existence and stability of a 
$p/q$ frequency-locked state. We also introduce various types of non-smooth bifurcations observed in the E-I pair. The bifurcation structure of the $p/q$ frequency-locked regions obtained through our analytical method is described in section \ref{sec:Bifurcations_And_multistability}. The summary of our work and future directions are described in conclusion.

\section{Event-driven maps}
\label{sec:Eqn_of_motion}
We consider the Leaky Integrate-and-fire (LIF) model of a neuron in our study. In this model, the voltage of an uncoupled LIF neuron rises monotonically to a threshold as a steady current flows into it. At the threshold, the neuron discharges instantaneously and the voltage drops to a reset value giving rise to a voltage spike. From its reset the voltage begins to rise once again towards the threshold and the process repeats. For a neuron synaptically coupled to others there is an additional current flowing into the neuron, through the synapses. We consider chemical synapses where the synaptic current is an explicit function of time taking the form of a pulse.  Until a pulse-coupled neuron reaches the threshold, its voltage evolves according to the differential equation
\begin{equation}
    \frac{dx_{i}}{dt}=a-x_{i}+g_{i}E_{i}(t),
    \label{Eqn:LIFmodeleqn}
\end{equation}
where $x_i$ is the voltage of neuron $i$ with $i=1,2$ for excitatory and inhibitory neurons respectively. The reset condition at the threshold crossing is expressed as
\begin{equation*}
    x_{i}\geq x_{th},\quad  x_{i}\leftarrow x_{reset}
    \label{Eqn:reset}
\end{equation*}
where $x_{th}$ is the threshold voltage and $x_{reset}$ is the reset voltage. We set $x_{th}=1$ and $x_{reset}=0$ in our work. To obtain a spike, the threshold voltage should be lower than the value of $a$. We set $a=1.3$ for both neurons. The free running frequency of a LIF neuron is uniquely determined by the parameter $a$ and is given by $\ln{\Big(\frac{a}{a-1}\Big)}$. Thus, the uncoupled neurons (one excitatory and another inhibitory), possess \textit{equal spiking frequencies} in our study and are consequently identical.  A spike in neuron $i$ initiates current injection into neuron $j \neq i$ synaptically coupled to $i$. The term $g_iE_i$ in Eq.(\ref{Eqn:LIFmodeleqn}) is the synaptic current into neuron $i$. Strength of the synapse formed by neuron $2$ (neuron $1$) onto neuron $1$ (neuron $2$) is denoted by $g_{1}(g_2)$. In this work, we  focus on a neuronal pair where one is excitatory (E) and the other is inhibitory (I). We will regard neuron $1$ as excitatory and neuron $2$ as inhibitory. Since neuron $1$ receives an inhibitory current $g_1<0$. Likewise, $g_{2}>0$. In our work, the \textit{magnitude of $g_1$ and $g_2$ are equal}. The time evolution of the synaptic current due to a single spike is often modeled as an alpha pulse given by $f(u)=\alpha^{2}u e^{-\alpha u}$, where $u$ is the time elapsed since the last spike and  $\alpha$ is the inverse of synaptic time constant \cite{Sterratt2011}. The net current into neuron $i$ at time $t$ is then a sum over synaptic currents resulting from all spikes of neuron $j\neq i$ in the past. Let $\{t_n\}$, with $ n=-\infty \cdots,-2,-1,0,1,2,\cdots,+\infty$ be the spike times of the two-neuron network arranged in an increasing order of time. Let us identify the neuron firing the $n^{th}$ spike by $s(n)$, with $s(n)=1,2$. We may now express net current into neuron $i$ as
\begin{equation}
    g_{i}E_{i}(t)={g_{i}}\sum_{n}C_{s(n),i}\theta(t-t_{n})f(t-t_{n}) ,
    \label{Eqn:Netcurrent}
\end{equation}
where, $C_{s(n),i} = 1$ if $s(n)\neq i$ and $C_{s(n),i}=0$ otherwise while $\theta(x)$ is Heaviside function. The sum in Eq.(\ref{Eqn:Netcurrent}) requires information of all spike timings in the past. We follow \cite{Olmi2010} wherein this requirement of an infinite summation is removed by converting Eq.(\ref{Eqn:Netcurrent}) into a second order differential equation for $E_{i}(t)$ given by
\begin{align}
  \frac{d^2E_{i}(t)}{dt^{2}}+2\alpha\frac{dE_{i}(t)}{dt^{2}}
   +\alpha^{2}E_{i}(t)={\alpha^{2}}\sum_{n|t_n<t} C_{s(n),i}\delta(t-t_{n}) ,
    \label{Eqn:diff.eqn.forE(t)}
\end{align}
following which Eqs.(\ref{Eqn:LIFmodeleqn}) and (\ref{Eqn:diff.eqn.forE(t)}) are transformed into discrete time maps as follows. Voltage of the neuron responsible for $n^{th}$ network spike is reset to zero and Eqs. (\ref{Eqn:LIFmodeleqn}) and (\ref{Eqn:diff.eqn.forE(t)}) are integrated from time $t^{+}_{n}$  to  $t^{-}_{n+1}$ where $t^{+}_{n}$ and $t^{-}_{n+1}$ are the times immediately after the $n^{th}$ spike and just before the $(n+1)^{th}$ spike respectively. In the event-driven maps that follow a superscript \lq$-$\rq  (\lq$+$\rq) on a variable denotes their value just before (after) a network spike.
The resulting event-driven maps are
\begin{equation}
    x^{-}_{i}(n+1)=x^{+}_{i}(n)e^{-\tau(n)}+a(1-e^{-\tau(n)})+g_{i}H_{i}(n) ,
\label{Eqn:x_minus_map}
\end{equation}
\begin{equation}
    E^{-}_{i}(n+1)=\big(E^{+}_{i}(n)+Q_{i}^{+}(n)\tau(n)\big)e^{-\alpha\tau(n)},
   \label{Eqn:E_minus_map}
    \end{equation} 
\begin{equation}
    Q^{-}_{i}(n+1)=Q^{+}_{i}(n)e^{-\alpha\tau(n)},
    \label{Eqn:Q_minus_map}
\end{equation}
with $+$ and $-$ variables related as
\begin{equation}
    x^{+}_{i}(n+1)=C_{s(n+1),i}x^{-}_{i}(n+1),
    \label{Eqn:x_plus_map}
\end{equation}

\begin{equation}
    E^{+}_{i}(n+1)=E^{-}_{i}(n+1),
    \label{Eqn:E_plus_map}
\end{equation}

\begin{equation}
    Q^{+}_{i}(n+1)=Q^{-}_{i}(n+1)+\alpha^2C_{s(n+1),i} ,
\label{Eqn:Q_plus_map}
\end{equation}
and
\begin{align}
\nonumber
    H_{i}(n)=\frac{e^{-\tau(n)}-e^{-\alpha\tau(n)}}{\alpha-1}\Big(E^{+}_{i}(n)
    +\frac{Q^{+}_{i}(n)}{\alpha-1}\Big)\\
    -\frac{Q^{+}_{i}(n)e^{-\alpha\tau(n)}\tau(n)}{\alpha-1}.
\label{Eqn:Hmap}    
\end{align}
In the above equations, $Q_i=\alpha E_i+\frac{dE_i}{dt}$ and $\tau(n)=t_{n+1}-t_{n}$ is the interspike interval in the network. To complete the maps in Eqs.(\ref{Eqn:x_minus_map})-(\ref{Eqn:Hmap}) we need to determine $\tau(n)$. This is done through calculation of the time taken $\tau_1$ and $\tau_2$ for neuron $1$ and $2$ respectively to reach the threshold by setting $x_{i}^{-}(n+1)$ in Eq.(\ref{Eqn:x_minus_map}) equal to one for both neurons. Expressing the function on the right hand side of Eq.(\ref{Eqn:x_minus_map}) as $\mathcal{F}_{i}(x,E,Q,\tau)$, solutions of 
\begin{equation}
x_{i}^-(n+1)=\mathcal{F}_{i}(x^{+}_{i}(n),E^{+}_{i}(n),Q^{+}_{i}(n),\tau_{i}(n))=1
    \label{Eqn:eqnfortau_i}
\end{equation}
will yield $\tau_1(n)$ and $\tau_2(n)$. Eq.(\ref{Eqn:eqnfortau_i}) can have multiple $\tau_1>0$ but unique $\tau_2>0$ solutions, as we explain next. The above maps are obtained by integrating between two spikes. However, if we disregard resetting, the integration could be performed from $t_n$ to any time in future. The result will give an extended voltage trajectory $x_i^{-}(u)=\mathcal{F}_{i}(x^{+}_{i}(n),E^{+}_{i}(n),Q^{+}_{i}(n),u)$ with $u=t-t_n^{+}$, obtained by replacing $(n+1)$ on the left and $\tau(n)$ on the right hand side of Eq.(\ref{Eqn:x_minus_map}). A unique solution of $x_2^{-}(u)=1$ is always obtained for neuron $2$ that receives an excitatory current. This is because voltage of this neuron, being driven towards the threshold by both the external current and the excitatory synaptic current, rises monotonically. Hence $x^{-}_2(u)=1$ for a single value of $u=\tau_2^*(n)$. In other words Eq. (\ref{Eqn:eqnfortau_i}) has a unique solution for neuron $2$.  Neuron $1$ receives an inhibitory alpha pulse which tends to push the neuron's voltage away from the threshold. This effect of inhibitory pulse competes against the constant external current. The net effect is such that $x_1^{-}(u)=1$ may be satisfied at multiple values of  $u$ which correspond to multiple solutions for $\tau_1(n)$ in Eq.(\ref{Eqn:eqnfortau_i}). However, the neuron will reset at the first threshold crossing. This feature is overlooked in some of the earlier analytical works with inhibitory neurons \cite{Coombes1999,van_vreeswijk_1994}. The first threshold crossing  
    $\tau_1^*(n)=\min\{\tau_1(n)\mid \tau_1(n)\in \mathbb{R}^{+} \hspace{0.1cm} \textrm{where} \hspace{0.1cm} \tau_1(n) \hspace{0.1cm}\textrm{is a solution of Eq.(\ref{Eqn:eqnfortau_i})}\}$. Once $\tau_1^*(n)$ and $\tau_2^*(n)$ are known then $\tau(n)=\min\{\tau_1^*(n),\tau_2^*(n)\}$ in Eqs.(\ref{Eqn:x_minus_map})-(\ref{Eqn:Hmap}).
\par
\begin{figure*}[t]
    \centering
    \includegraphics[width=0.8\textwidth]{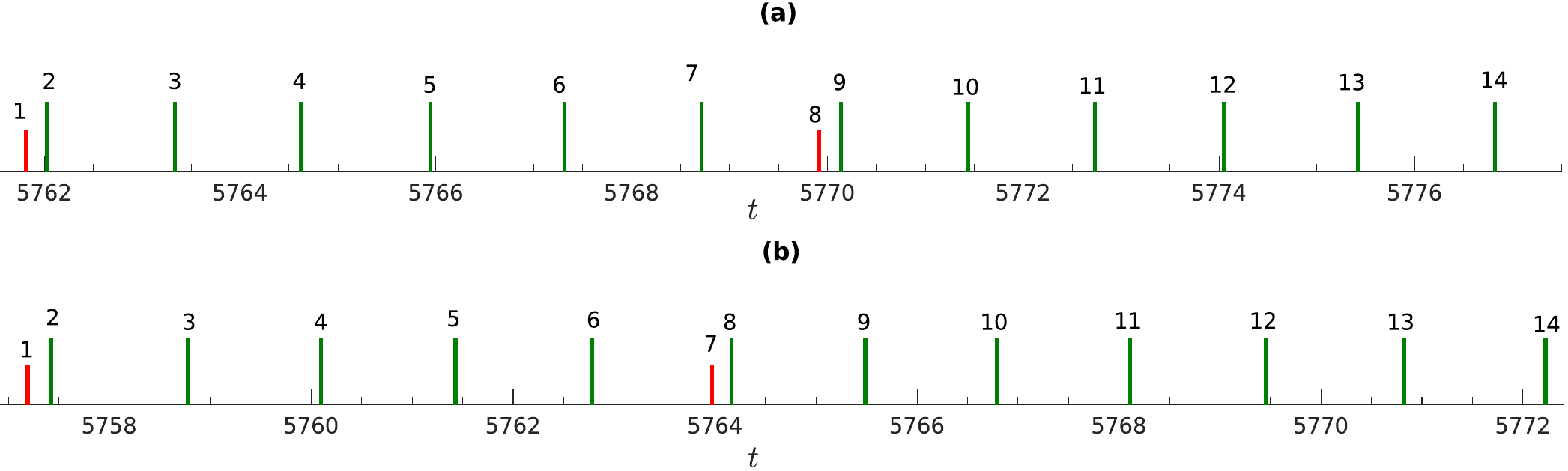}
    \caption{
    (Color online) Spike pattern of excitatory (inhibitory) neurons for (a) Two cycles of $1/6$  frequency locking with spike sequence $\{1,2^6\}$ for $g=0.404238$ and $\alpha=0.526$ and 
(b) a single cycle of $2/12$ frequency locking with spike sequence $\{1,2^5,1,2^7\}$ is shown for $g=0.40374$ and  $\alpha=0.374$. A shorter spike (in red) represents the spike of an excitatory neuron while a taller spike (in green) represents the spike of an inhibitory neuron. The network spike number in the sequence is labeled on top of each spike.}
    \label{fig:frequencylocking_1by6_and2by12}
\end{figure*}
In our work, we have employed event-driven maps to carry out numerical simulations. We have also formulated an analytical approach, based on these maps, to determine regions of frequency locking in $(g,\alpha)$ parameter space. We find periodic as well as non-periodic spike sequences in the E-I neuron pair. Let us define an average firing rate of neuron $i$ as  $F_i=\lim_{\substack{t\rightarrow \infty}}\frac{N_i}{t}$ where $N_i$ is the number of spikes in time interval $t$. The rotation number in terms of $F_i$ can be written as $\rho=\frac{F_1}{F_2}$. Within a single period of repeating sequence of interspike intervals in the network, if there are $p$ spikes of neuron $1$ and $q$ spikes of neuron $2$, there is $p/q$ frequency locking. We will identify both irreducible $p/q$ and reducible $np/nq$ frequency locking ratios where $p$ and $q$ are relatively prime.
Rotation number is expressed as the simplest equivalent fraction and thus the same for $p/q$ and $np/nq$. However $p/q$ and $np/nq$ have distinct set of $(p+q)$ and $n(p+q)$ interspike intervals.
As an example of this we show $1/6$ and $2/12$ frequency-locked states in Fig.\ref{fig:frequencylocking_1by6_and2by12}.
\par
We will represent periodic spike sequences of a $p/q$ frequency-locked state in two distinct manners, chosen based on what notation is most convenient. One of the representations will be given by a repeating spike sequence  $\{s(0),s(1),s(2),\cdots,s(p+q-1)\}$. In this representation, a spike sequence of $2/5$ is $\{1,2,2,1,2,2,2\}$. In the other representation, this sequence will be expressed as $\{1,2^{2},1,2^{3}\}$.
\par
In our identical E-I pair, an excitatory neuron cannot produce two consecutive network spikes. Consider the scenario when the excitatory neuron has emitted $n^{th}$ network spike and reset its voltage to zero, i.e. $x_{1}^+(n)=0$. At this juncture, the voltage of the inhibitory neuron, $x_{2}^+(n)$, will exceed that of the excitatory neuron between the reset and threshold. Considering that in a E-I neuron pair, the excitatory neuron receives inhibitory current while the inhibitory neuron receives excitatory current, and aside from this distinction both neurons are identical, the inhibitory neuron will progress more rapidly towards the threshold, leading it to fire the $(n+1)^{th}$ spike. This observation implies that the $(n+1)^{th}$ spike cannot emanate from the excitatory neuron that last reset. Consequently, we can infer that $p<q$.
\par
Multiple arrangements of excitatory-inhibitory spikes are possible for most $p/q$ ratios. For instance, the ratio $2/5$ can manifest through either $\{1,2^{2},1,2^{3}\}$ or $\{1,2^{1},1,2^{4}\}$. However, specific frequency-locking patterns like $1/q$ and $q/(q+1)$ can only accommodate a single spike sequence. In the case of $1/q$ this uniqueness is evident, while for $q/(q+1)$ it arises from the stipulation that excitatory neuron cannot consecutively emit two network spikes. For other $p/q$ frequency lockings, our work reveals (in Sec.\ref{sec:Bifurcations_And_multistability}) that only a few of the potential $p/q$ spike sequences are realized.

\section{Frequency locking: Simulation Results}
\label{sec:simulation_results}
We explore the dynamics of an \textit{identical} E-I neuron pair, modeled as LIF oscillators, across $(g,\alpha)$ space, through iterations of the event-driven maps Eqs.(\ref{Eqn:x_minus_map})-(\ref{Eqn:Hmap}). The same initial condition is used for all simulations. Typically, the first $1000-3000$ spikes are disregarded for transients to die out, where this duration depends on the values of $\alpha$ and $g$. Following this we (i) compute $\rho$ over a span of approximately $500$ spikes, (ii) find the frequency locking ratio, and (iii) examine the spike sequence obtained. The values of $\alpha$ and $g$ are varied from $0<\alpha\leq 30$ and $0<g\leq1.25$ respectively. These parameters are systematically adjusted to observe the changes in the synchronization behavior of neuron pair.
\begin{figure}[t]
    \centering
    \includegraphics[width=0.45\textwidth]{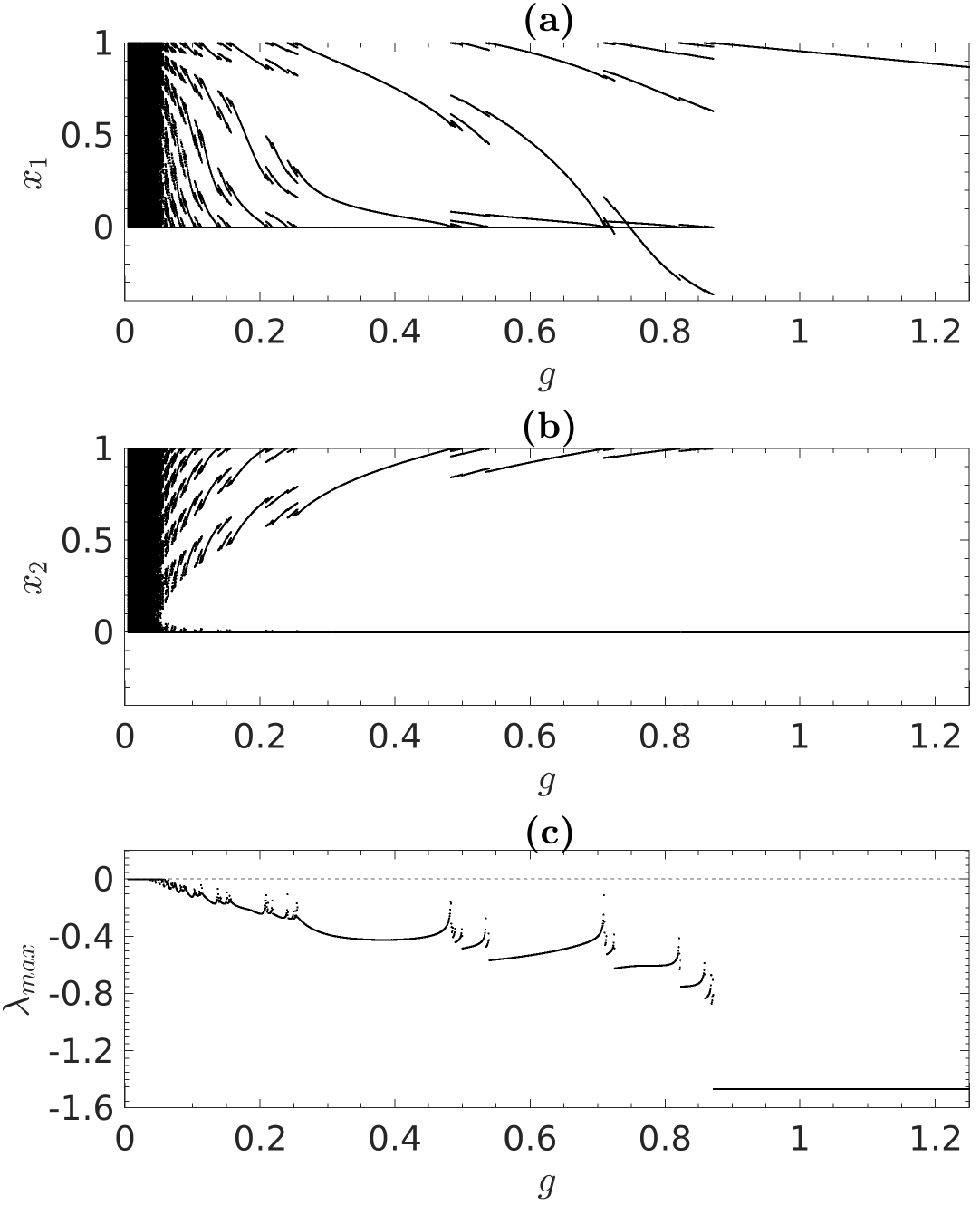}
    \caption{Voltage values immediately after a network spike for (a) excitatory neuron (b) inhibitory neuron, as a function of synaptic strength $(g)$.  (c) Maximal Lyapunov exponent $(\lambda_{max})$ vs $g$. Simulation parameters: $\alpha=15,  
 |g_1|=|g_2|=g$ .}
    \label{fig:bifurcation diagram alpha=15}
\end{figure}
\par
The E-I pair exhibits either periodic or quasiperiodic dynamics depending upon values of $g$ and $\alpha$. A representative set of dynamics is displayed in Fig.\ref{fig:bifurcation diagram alpha=15} (a)-(c) through a pair of bifurcation diagrams and Lyapunov exponents by varying $g$ for a specific $\alpha$. The pair of diagrams displays the voltage values of the excitatory and inhibitory neurons whenever a spike occurs in either of the two neurons. The voltage value of the neuron that spiked is shown in the diagram after resetting its value to zero while that of the other one is a non-zero value lying below $x_{th}$. For the case of periodic dynamics if the number of distinct non-zero voltage values of neuron $2$ and neuron $1$ are $p$ and $q$ respectively then they are $p/q$ frequency-locked. For example, at $g=0.4$ neuron $1$  and neuron $2$   have  two and one distinct non-zero voltage values respectively yielding  $1/2$ frequency-locked state. A number of different $p/q$ intervals of different sizes exist. For large $g$ the excitatory neuron ceases to fire while the inhibitory neuron fires periodically. As a result, the rotation number is $0/1$. The firing death of the excitatory neuron is seen in the absence of a branch at $x=0$ in the bifurcation diagram for the excitatory neuron. Such suppression of firing of a neuron is known to occur if inhibitory neurons are present in the network \cite{Chow1998,AnguloGarcia2017}. For small $g$ no periodicity is evident in the bifurcation diagram. We further explore the character of the dynamics over these $g$'s by computing the maximal lyapunov exponent (see Fig. 
 \ref{fig:bifurcation diagram alpha=15}(c)) $\lambda_{max}$. Over this aperiodic interval, $\lambda_{max}\approx 0$ indicating that the dynamics is quasiperiodic. In contrast and as expected, $\lambda_{max}<0$ over the frequency-locked regions.
 \begin{figure}[t]
     \centering
    \includegraphics[width=0.5\textwidth]{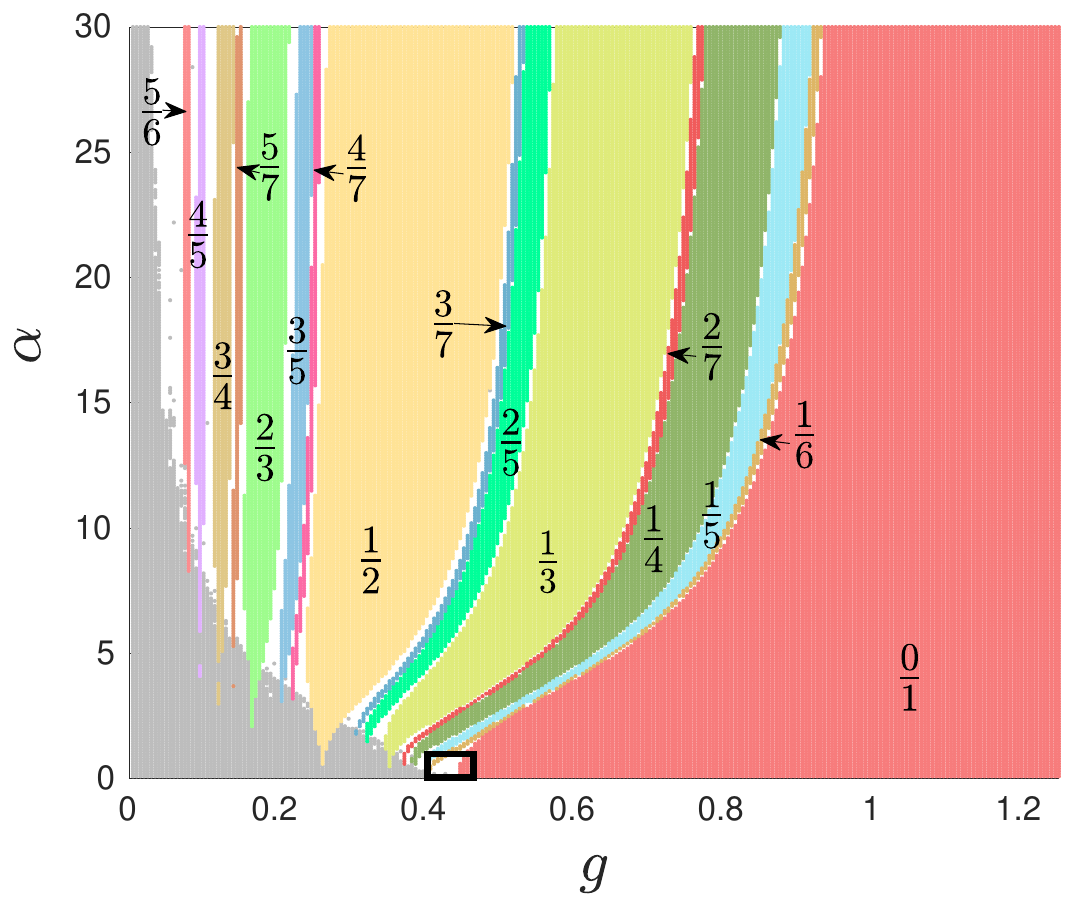}
     \caption{ (Color online) Dominant sixteen frequency-locked $p/q$ regions and quasiperiodic region (shown in grey). The small boxed region is resolved further in Fig.\ref{fig:pspace_higher_resolution}. }  \label{fig:pspace_low_Resolution_simulations}
 \end{figure}
 \par
The dynamics of the E-I neuronal pair across $(g,\alpha)$ parameter space obtained from simulations is represented in Fig.\ref{fig:pspace_low_Resolution_simulations}. Simple $1/1$ frequency locking is absent.
The space is dominated by numerous high-order frequency-locked regions descending from large to small values of $\alpha$. These are reminiscent of Arnold tongues in forced and non-identical coupled oscillators,
and we will refer to these as \lq frequency-locked tongues' or merely \lq tongues\rq. In the only other reported work \cite{Bressloff1998}, of high-order synchronization in coupled \textit{identical} oscillators (specifically in an I-I pair) only a single occurrence ($1/2$ locking) was observed. However, our work with an identical E-I neuron pair reveals a far richer structure of high-order synchronization phenomena, with numerous $p/q$ locking regions across the system’s parameter space. Our findings of multiple high-order synchronization regions suggest that the dynamics in E-I neuron pairs are far more complex than previously understood dynamics of E-E or I-I pairs. Quasiperiodicity dominates over a portion of parameter space, particularly at small $\alpha$ and small $g$. Firing death is the largest region with periodic dynamics. The width of various frequency-locked regions increases with $\alpha$. At large $\alpha's$ the dynamics of the pair tends to become independent of $\alpha$. All the tongues shown in Fig.\ref{fig:pspace_low_Resolution_simulations} belong to irreducible $p/q$ ratios where $p$ and $q$ are relatively prime. We will refer to these as irreducible tongues. 
\par
\begin{figure*}[t]
    \centering
\includegraphics[width=\textwidth]{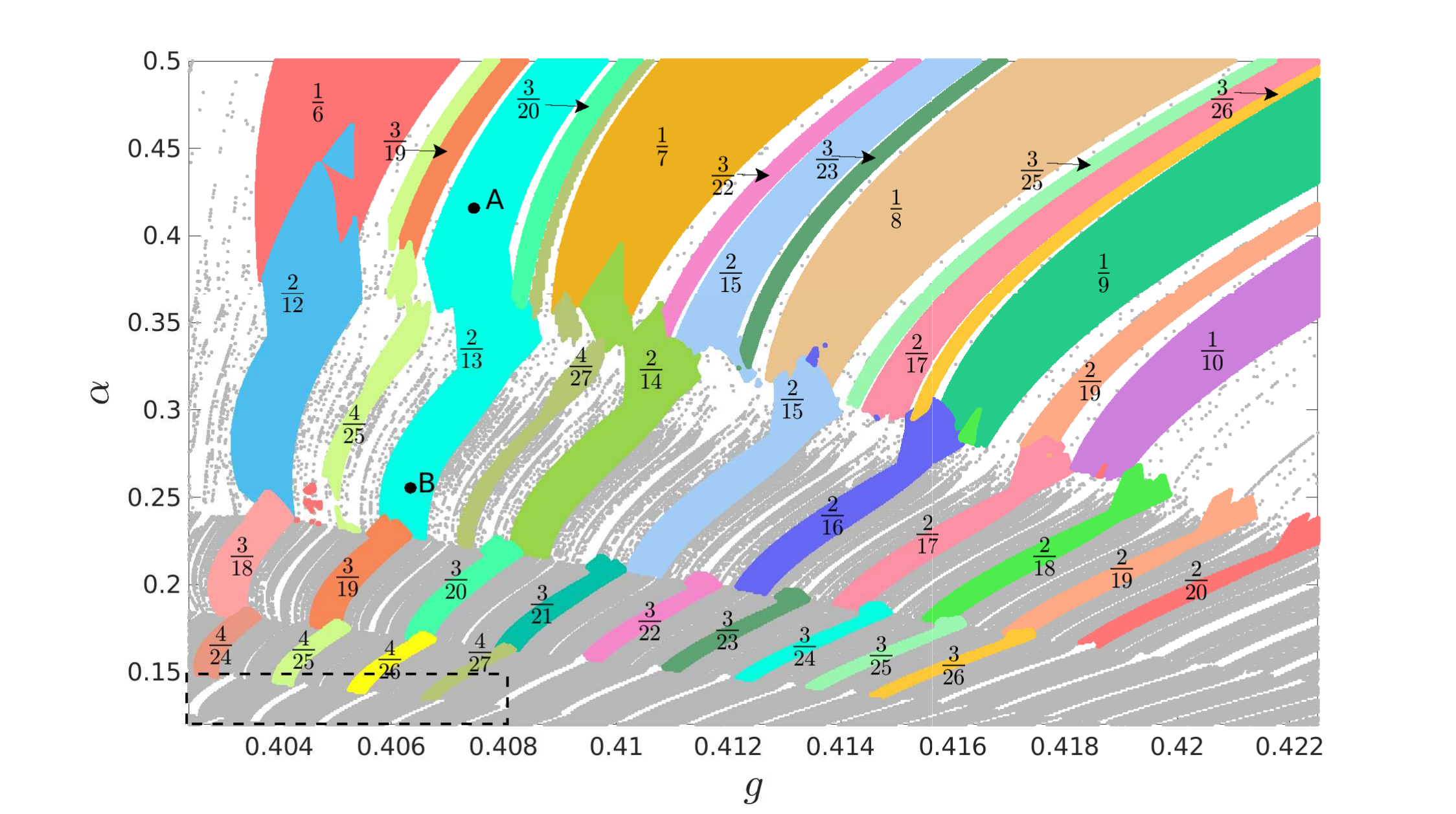}
    \caption{(Color online) Magnification of boxed region of Fig.\ref{fig:pspace_low_Resolution_simulations}. A few dominant irreducible $p/q$ and reducible $np/nq$ regions are shown. Each color represents a distinct frequency-locked region and is labeled. For every irreducible $p/q$ tongue, a sequence of $np/nq$ islands exists. For instance, the following sequence can be seen $1/6,2/12,3/18,4/24$. Other sequences are shown up to varying levels, $n$, restricted by resolution and space in the figure. The boxed region is resolved further in Fig.\ref{fig:np_by_nq_islands} and displays more such sequences.
    Higher $n$ islands originate at lower values of $\alpha$. All the tongues and islands appear to end abruptly as they descend and some of them even reappear (e.g. $2/15,3/19,2/12$). Later on, the analytical method shows that multistability is responsible for this abruptness. For fixed $p$, a sequence of $p/q$ regions with increasing $q$ is arranged from lower to higher $g$ values. Points labeled as \lq A' and \lq B' inside the tongue $2/13$ have different spike sequences.}
\label{fig:pspace_higher_resolution}
\end{figure*}

 The boxed portion of Fig.\ref{fig:pspace_low_Resolution_simulations} at small $\alpha$ values, near the firing death boundary, is resolved further in Fig.\ref{fig:pspace_higher_resolution}, revealing the presence of a richer and more diverse frequency locking structure than visible in Fig.\ref{fig:pspace_low_Resolution_simulations}. In addition to the $p/q$ tongues that descend from large $\alpha$ values, we notice the presence of frequency-locked regions that exist as islands. These islands are formed from irreducible as well as reducible $np/nq$ ratios, with $p$ and $q$ being relatively prime. The irreducible tongues descending from large $\alpha's$ end abruptly and then reappear as islands at smaller $\alpha's$.
In Sec.\ref{sec:Bifurcations_And_multistability}, through an analytical approach, we establish that the irreducible $p/q$ islands are infact connected to the corresponding  $p/q$ tongues and also extend further down. The perceived disconnection stems from multistability at lower $\alpha$ values, where a given $p/q$ may not be attainable from the chosen initial condition. Conversely, analytical procedure will establish that the reducible $np/nq$ islands (such as $2/12$, $2/14$, etc.) in Fig.\ref{fig:pspace_higher_resolution} stand as distinct entities and emerge only at small $\alpha$ values. Therefore, frequency-locked regions defined by irreducible $p/q$ ratios are organized as tongues and those by reducible $np/nq$ ratios as islands.
\begin{figure*}[t]
    \includegraphics[width=\textwidth]{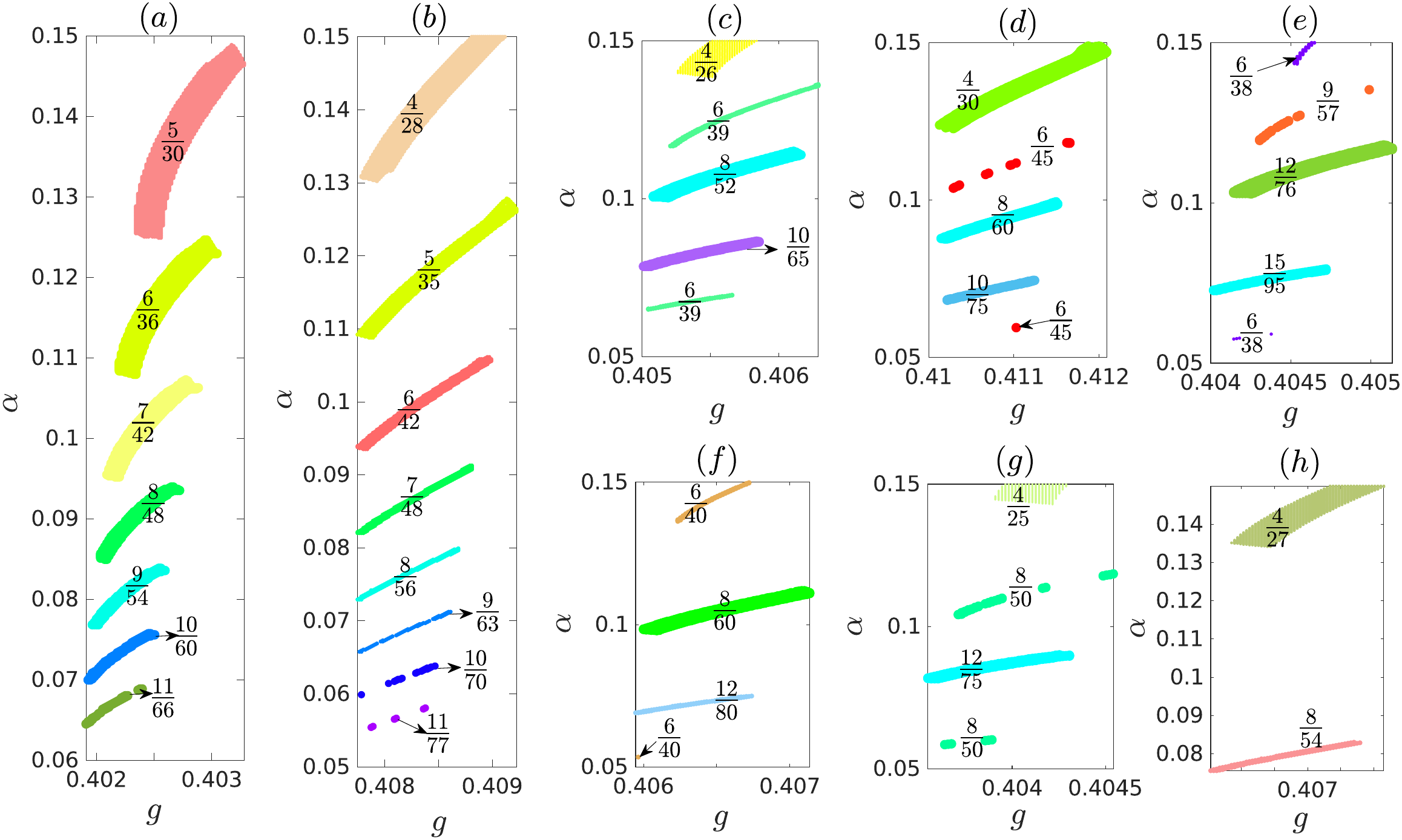} 
    \caption{(Color online) Few sequences of $np/nq$ islands in the boxed region of Fig.\ref{fig:pspace_higher_resolution} for (a) $p=1, q=6$, (b) $p=1$, $q=7$, (c) $p=2 ,q=13$, (d) $p=2, q=15$, (e) $p=3 ,q=19$, (f) $p=3, q=20$ (g) $p=4, q=25$ (h) $p=4, q=27$.}
\label{fig:np_by_nq_islands}
\end{figure*}
These islands, like the tongues, are also seen to end abruptly in Fig.\ref{fig:pspace_higher_resolution}. This abrupt ending will similarly be demonstrated, in Sec.\ref{sec:Bifurcations_And_multistability}, to arise from multistability. A $np/nq$ island has a periodicity $n(p+q)$. In Fig.\ref{fig:pspace_higher_resolution}, we observe that for a fixed $p/q$, various
$np/nq$ islands are created in ascending order of their periodicity, as $\alpha$ is lowered. For example, corresponding to $p=1$ and $q=6$ the islands  $2/12,3/18,4/24$ emerge sequentially on lowering $\alpha$.
In the figure, these sequence of islands appears to be disjoint. In Sec.\ref{sec:Bifurcations_And_multistability}, we present evidence that such family of islands have mutual overlaps. In Fig.\ref{fig:np_by_nq_islands} we further resolve small $\alpha$ region of Fig.\ref{fig:pspace_higher_resolution}
to find out whether the $np/nq$ sequence of islands appears only for some $p/q$ or it is generic. In the figure, we show sequence of islands corresponding to $p/q's$, with $p$ ranging from $1$ to $4$ for few chosen $q's$. For $p=1$ we resolved islands upto $n=11$ while for higher values of $p$ fewer islands are resolved. While searching for islands we could resolve islands with greater values of $n$ by increasing the resolution. At the same time, we could also begin observing islands corresponding to larger $p$ values which were otherwise invisible at lower resolutions. These observations lead us to believe that sequence of islands exists for all $p/q's$ at small $\alpha$ values, close to the firing death boundary, and for every $n$.
\par 
\begin{figure}[t]
    \centering
    \includegraphics[width=0.5\textwidth]{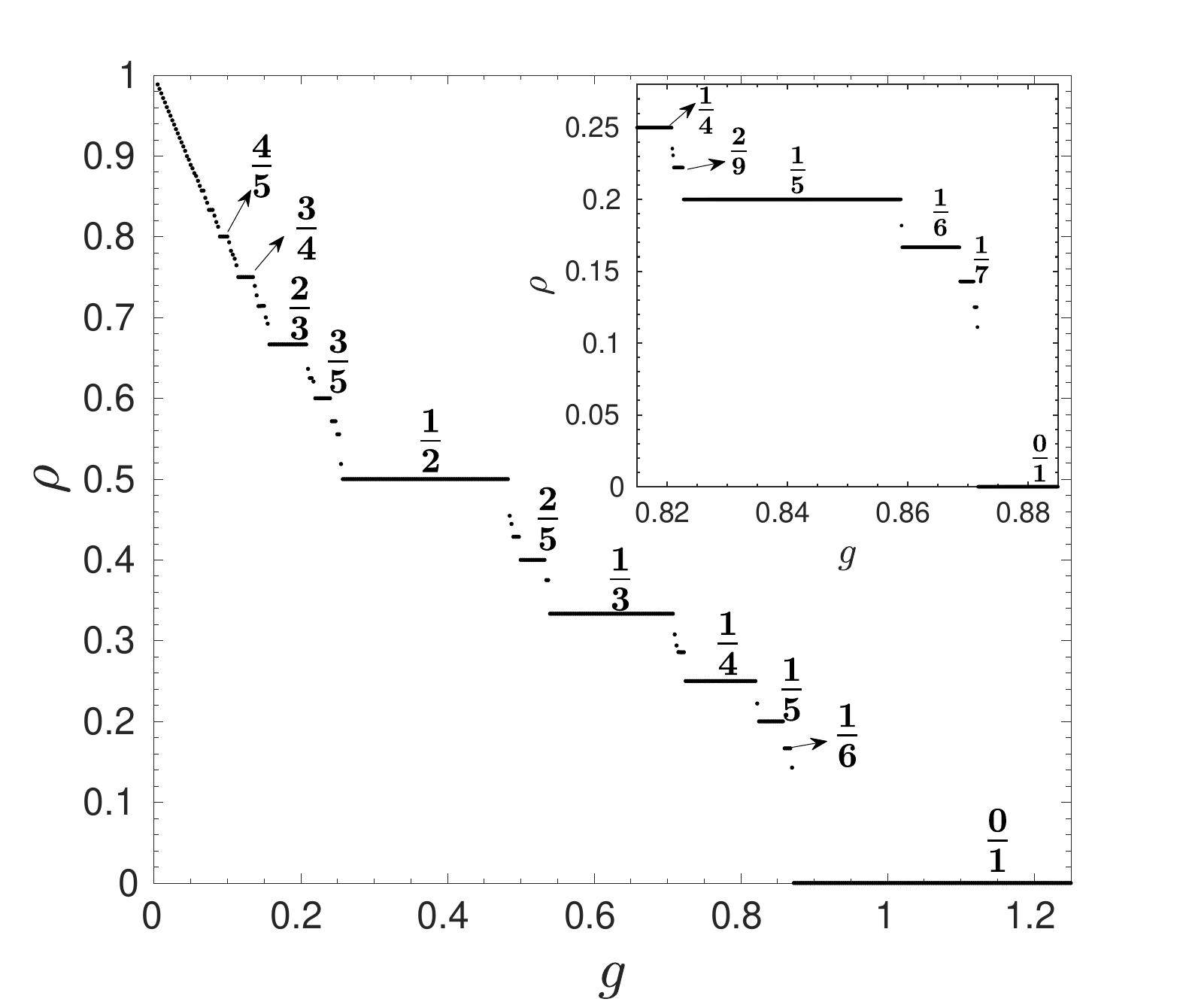}
    \caption{Devil staircase structure for $\alpha=15$}
    \label{fig:devil_staircase}
\end{figure}
\begin{figure*}[t!]
    \centering
    \includegraphics[width=0.9\textwidth]{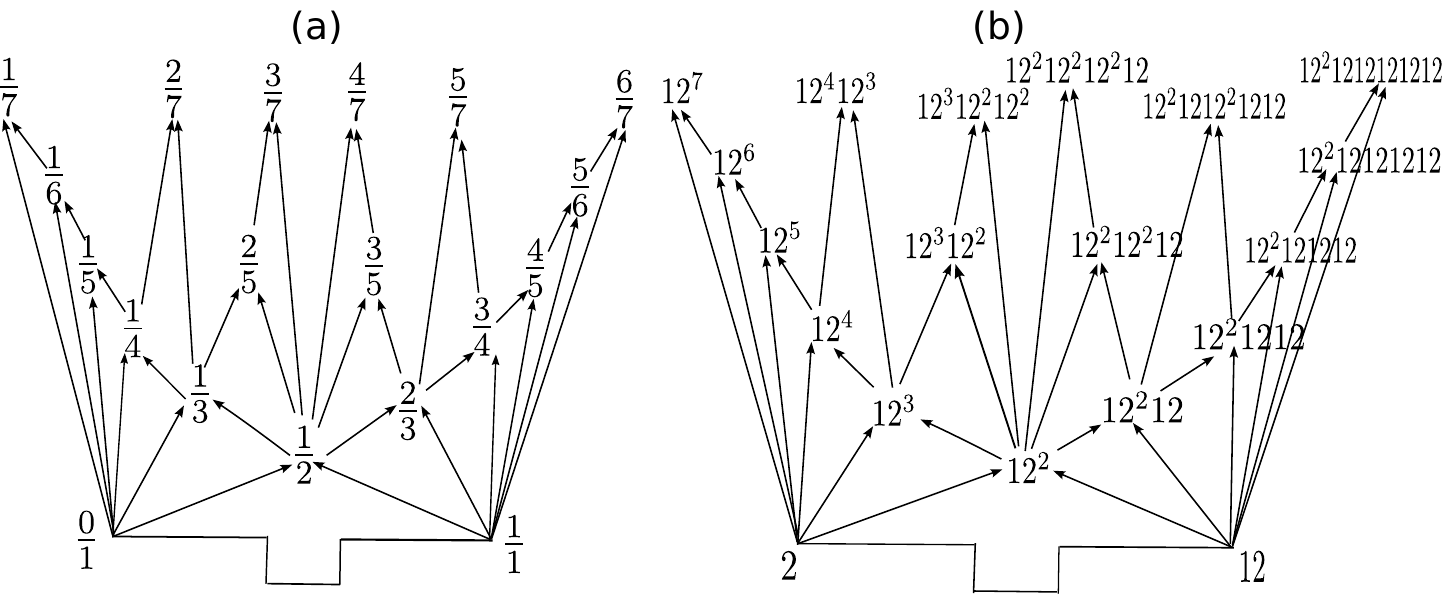}
    \caption{Farey tree arrangement of (a) frequency locking Farey tree and (b) spike sequence Farey tree. Curly brackets in the spike sequences are removed for convenience. $1/1$ frequency locking shown in the above figure does not occur for $g\neq 0$ but will occur if $g=0$. For every frequency locking ratio in its Farey tree corresponding spike sequence exists at the same level in the spike sequence Farey tree.}
    \label{fig:farey_tree}
\end{figure*}

At higher $\alpha$ values, in Fig.\ref{fig:pspace_low_Resolution_simulations}, the $p/q$ tongues are organized along $g$ following the well-known Farey tree arrangement \cite{hilborn2000chaos}. Between any two irreducible frequency locking ratios $p/q$ and $p'/q'$ another frequency-locking ratio $(p+p')/(q+q')$ exists. This is visible in Fig.\ref{fig:pspace_low_Resolution_simulations} but becomes more evident in Fig.\ref{fig:devil_staircase} illustrating $\rho$ vs. $g$ for a specific $\alpha$. The figure also shows the presence of a devil staircase. We also examined the spike sequences across those $p/q$ tongues where multiple spike sequences are possible. In such cases we find the same spike sequence may not exist across the entire tongue. For instance, in the case of $2/13$ at points \lq A' and \lq B', in Fig.\ref{fig:pspace_higher_resolution}, we observe the spike sequences $\{1,2^6,1,2^7\}$ and $\{1,2^5,1,2^8\}$ respectively.
\par
At higher $\alpha$  values where the Farey tree arrangement of frequency locking ratios is observed, there also exists a Farey arrangement of spike sequences. This is similar to the Farey arrangement of symbolic sequences in one-dimensional piece-wise continuous maps \cite{Granados2017} and discontinuous periodic forcing of a spiking model\cite{Granados2015}. A $p/q$ and $p'/q'$ are Farey parents of $P/Q$ if and only if they satisfy the following conditions
\begin{enumerate}
    \item $p/q<p'/q'$
    \item $p'q-q'p=1$
    \item $P/Q=(p+p')/(q+q')$
\end{enumerate}
 In a Farey arrangement of spike sequences, the spike sequence of $P/Q$ is given by the concatenation of spike sequences of $p/q$ and $p'/q'$. For example, $1/6$ and $1/7$ are Farey parents of $2/13$. Spike sequences of $1/6$ and $1/7$ are $\{1,2^6\}$ and $\{1,2^7\}$ respectively.
 In our simulations, at large $\alpha$ values (for instance at A in Fig.\ref{fig:pspace_higher_resolution}) $2/13$ is realized through the spike sequence $\{1,2^{6},1,2^{7}\}$ which is a concatenation of the spike sequences of its Farey parents. Fig.\ref{fig:farey_tree} elaborates the Farey arrangement of spike sequences corresponding to other $p/q$ frequency-locked states. In the context of neuronal dynamics, such an arrangement of spike sequences has not been reported earlier. At lower $\alpha's$ the Farey arrangement of rotation numbers and sequences gets broken. This happens due to multistability at these $\alpha's$. Additionally, the spike sequences at lower $\alpha$ values need not be same as the ones at higher $\alpha$.

\section{Frequency-locking: Analytical Formulation}
\label{sec:pbyq_freq_locked_analytical}

\subsection{Existence of solutions}
\label{subsec:existence_of_solutions}
In this section, we formulate an approach to determine the region of existence, in parameter space, of a \textit{specified periodic spike sequence} $\{s(0),s(1),s(2),\cdots,s(p+q-1)\}$ of $p$ spikes of neuron $1$ and $q$ spikes of neuron $2$ in a $p/q$ frequency-locked state. In general, our approach holds for any combination of a pair of excitatory and inhibitory neurons. However, the primary focus of this work is on an E-I pair. Our approach employs event-driven maps. 
\\ \\ 
\textbf{Step 1:}  We begin by scheduling a spike sequence $\{s(-\infty)...s(-1), s(0), s(1), s(2),...,s(n),s(n+1),....s(\infty)\}$. The sequence is arbitrary, subject to the restriction that two successive spikes cannot come from excitatory neuron (see Sec. \ref{sec:Eqn_of_motion} ). Constraint of periodicity is applied in the next step. Furthermore, we assume arbitrary values for $\tau(n)$ instead of fixing these using Eq.(\ref{Eqn:eqnfortau_i}). Implying that we forcibly reset neuron $s(n+1)$ after an interspike interval $\tau(n)$ without considering its voltage at that time. As such, the neuron could be reset even before it reaches the threshold or get reset after it reaches the threshold. In the former case we get a shortened voltage trajectory and in the latter case an extended voltage trajectory in the underlying dynamics. We will impose Eq.(\ref{Eqn:eqnfortau_i}) in step 3 to ensure that reset occurs at the threshold.

\begin{figure*}[t]
    \centering
    \includegraphics[scale=0.25]{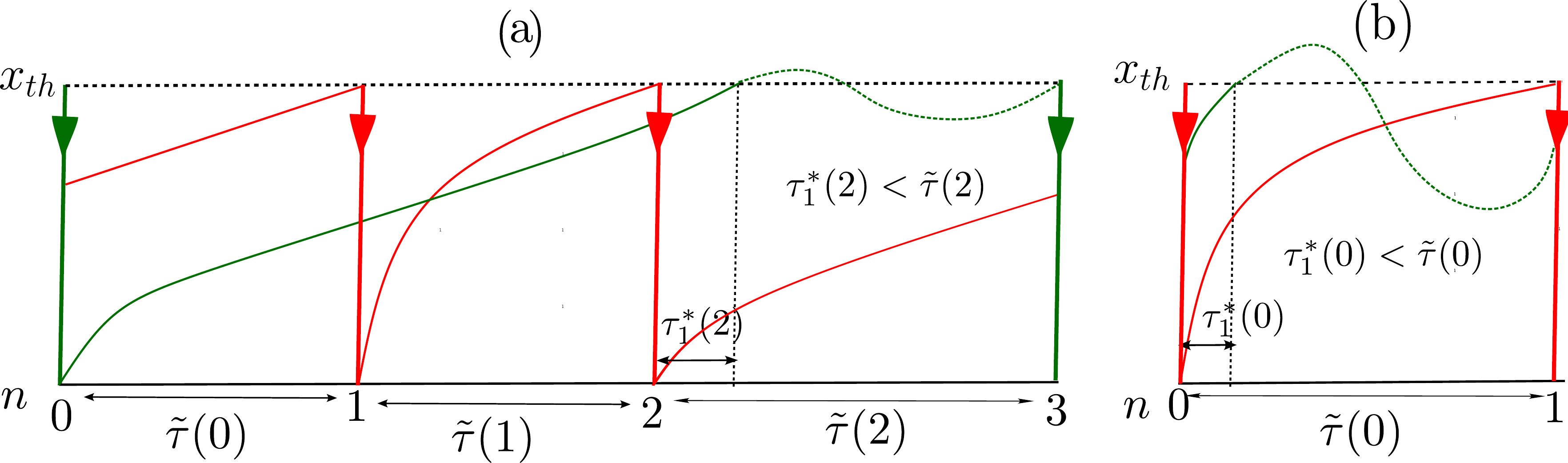}
    \caption{(Color online) Schematic figures of  extended and actual voltage trajectories of excitatory (neuron $1$) and inhibitory (neuron $2$) are illustrated for two cases:
    (a) for $1/2$ frequency-locked solution with periodic spike sequence $\{1,2,2\}$ and for (b) $0/1$ firing death solution with periodic spike sequence $\{2\}$. Actual and extended voltage trajectories are shown in solid and dotted lines respectively. Excitatory and inhibitory neuron trajectories are shown in green and red respectively. The solid line with downward arrow represents the spike of a neuron. The spike numbers are marked with a numeral at the bottom. In (a) between spike number $2$ and $3$, first threshold crossing of excitatory neuron, at $\tau_1^*(2)$, occurs earlier than its threshold crossing $\tilde{\tau}(2)$ obtained through Eqs.(\ref{Eqn:tau_eqns_H(tau)}), violating Condition $1$. In (b) $1^{st}$ threshold crossing of neuron $1$ occurs before the crossing  $\tilde{\tau}(0)$ of neuron $2$, violating Condition $2$.}
    \label{fig:Conditions1and2}
\end{figure*}

We now construct a map expressing $x_{i},E_{i},Q_{i}$ \textit{after} $(n+1)$ scheduled spikes in terms of their values \textit{after} the $0^{th}$ spike (the initial state). We will first obtain $Q^{+}_{i}(n+1)$ followed by $E^{+}_{i}(n+1)$ and then $x^{+}_{i}(n+1)$. By repeatedly iterating Eq.(\ref{Eqn:Q_plus_map}) accompanied by Eq.(\ref{Eqn:Q_minus_map}) for $(n+1)$ times, $Q_{i}^{+}(n+1)$ can be written in terms of initial value $Q_{i}^{+}(0)$ and  $\{ \tau(j) \}_{j=0}^{n-1}$ as (see Appendix \ref{appendix:steps_For_calculations}  for details)
\begin{align}
\nonumber
    Q^{+}_{i}(n+1)&=Q^{+}_{i}(0)e^{-\alpha\Delta(0,n)}\\
    \nonumber
&+\alpha^{2}\Bigg[\sum\limits_{l=0}^{n}\Theta(l,n)C_{s(l),i}e^{-\alpha\Delta(l,n)}\Bigg]\\
    &+\alpha^{2}C_{s(n+1),i} \quad \text{,}
    \label{Eqn:Qi(n+1)_plus_in_terms_of_Qi(0)_plus}
\end{align}
where $\Delta(l,n)=\sum\limits_{j=l}^{n}\tau(j)$ for 
 $l=0,1,2,\cdots,n$, $l\leq n$ and 
 \begin{align}
\Theta(l, n) =
\begin{cases}
    0 & \text{if } l = 0 \text{ or } l > n \\
    1 & \text{if } 0 < l \leq n  \quad \text{.}
\end{cases}
\end{align}
Similarly using Eqs.(\ref{Eqn:E_minus_map}),(\ref{Eqn:Q_minus_map}),(\ref{Eqn:E_plus_map}) and (\ref{Eqn:Q_plus_map}) we get
\begin{align}
    E^{+}_{i}(n+1) &= \Big[E^{+}_{i}(0)+Q^{+}_{i}(0)\Delta(0,n)\Big]e^{-\alpha \Delta(0,n)} \nonumber +\\ &\alpha^{2}\Bigg[\sum\limits_{l=0}^{n}\Theta(l,n)C_{s(l),i}\Delta(l,n)e^{-\alpha\Delta(l,n)}\Bigg] \quad, \label{Eqn:Ei(n+1)_plus_in_terms_of_Ei(0)_plus}
\end{align}
which also depends on initial state and $\{\tau(j)\}_{j=0}^{n-1}$. In order to similarly obtain an expression of $x_{i}^{+}(n+1)$, for compactness let us re-express Eq.(\ref{Eqn:x_plus_map}) for $x^{+}_{i}(n+1)$ as
\begin{align}
\nonumber
    x_{i}^{+}(n+1)&=C_{s(n+1),i}\Big[x_{i}^{+}(n)e^{-\tau(n)}\\
    &+\zeta_{i}(\tau(n),E^{+}_{i}(n),Q^{+}_{i}(n))\Big]
   \label{Eqn:xi(n+1)plus} ,
\end{align}
where $\zeta_{i}(\tau(n),E^{+}_{i}(n),Q^{+}_{i}(n))=a(1-e^{-\tau(n)})+g_{i}H_{i}(n)$, and $H_{i}(n)$  given by Eq.(\ref{Eqn:Hmap}) depends on $E^{+}_{i}(n)$ and $Q^{+}_{i}(n)$.
To initialize the map $(n=0)$ we premultiply $x_{i}^{+}(0)$ by $C_{s(0),i}$. Doing this fixes the voltage  of the neuron firing the $0^{th}$ spike to zero while that of other neuron could be initialized to any value less than the threshold.
We can then write
\begin{align}
\nonumber
    x_{i}^{+}(1)&=C_{s(1),i}\Big[C_{s(0),i}x_{i}^{+}(0)e^{-\tau(0)}\\
    &+\zeta_{i}(\tau(0),E^{+}_{i}(0),Q^{+}_{i}(0))\Big]  .
    \label{Eqn:xi(1)plus}
\end{align}
Now iterating Eq.(\ref{Eqn:xi(n+1)plus}) accompanied by Eq.(\ref{Eqn:xi(1)plus}) for $(n+1)$ times we get (see Appendix \ref{appendix:steps_For_calculations})
\vspace{-32pt}
\begin{widetext}
\begin{align}
    x_{i}^{+}(n+1)  = x_{i}^{+}(0)e^{-\Delta(0,n)}P_{i}(0,n+1)+\Bigg[\sum\limits_{l=0}^{n}\Theta(l,n)
    \zeta_{i}(\tau(l-1), & E^{+}_{i}(l-1),Q^{+}_{i}(l-1))e^{-\Delta(l,n)}P_{i}(l,n+1)\Bigg]\nonumber  \\ 
     &+P_{i}(n+1,n+1)\zeta_{i}(\tau(n),E^{+}_{i}(n),Q^{+}_{i}(n))
     \label{Eqn:xi(n+1)_plus_in_terms_of_xi(0)_plus} ,
\end{align}
\end{widetext}
\vspace{-32pt}
where  $P_{i}(k,n)=\prod\limits_{l=k}^{n}C_{s(l),i}$. At this point, we have obtained a map given by Eqs.(\ref{Eqn:Qi(n+1)_plus_in_terms_of_Qi(0)_plus}),(\ref{Eqn:Ei(n+1)_plus_in_terms_of_Ei(0)_plus}) and (\ref{Eqn:xi(n+1)_plus_in_terms_of_xi(0)_plus})
connecting state of the system after the $(n+1)^{th}$ spike to its state after $0^{th}$ spike. These maps depend on $\{\tau(j)\}_{j=0}^{n-1}$ which are undetermined at this point.\\\\
\textbf{Step 2:} In a periodic spike sequence  $\{s(0),s(1),\cdots\,s(p+q-1)\}$ of a $p/q$ frequency-locked state, the state of the system soon after(before) $n^{th}$ network spike is the same as its state after(before) $(n+p+q)^{th}$ network spike. Hence, for $p/q$ frequency locking \romannumeral 1)
 $x^{+}_{i}(n+p+q)=x^{+}_{i}(n)=\tilde{x}^{+}_{i}(n)$,  \romannumeral 2) $Q^{+}_{i}(n+p+q)=Q^{+}_{i}(n)=\tilde{Q}^{+}_{i}(n)$, \romannumeral 3) $E^{+}_{i}(n+p+q)=E^{+}_{i}(n)=\tilde{E}^{+}_{i}(n)$, and \romannumeral 4) $\tau(n+p+q)=\tau(n)=\tilde{\tau}(n)$. We will now find expressions for $(p+q)$ distinct values (labeled by $m$) of $\tilde{x}^{+}_{i}(m),\tilde{E}^{+}_{i}(m)$ and $\tilde{Q}^{+}_{i}(m)$ in terms of undetermined $\{\tilde{\tau}(m)\}_{m=0}^{p+q-1}$. The $(p+q)$ distinct $\tilde{\tau}$'s will be later determined self consistently. Let us begin by finding expressions for $\tilde{x}^{+}_{i}(0)$, $\tilde{E}^{+}_{i}(0)$ and $\tilde{Q}^{+}_{i}(0)$ in terms of the undetermined $\{\tilde{\tau}(m)\}_{m=0}^{p+q-1}$. Putting $n+1=p+q$ in Eqs.(\ref{Eqn:Qi(n+1)_plus_in_terms_of_Qi(0)_plus}),(\ref{Eqn:Ei(n+1)_plus_in_terms_of_Ei(0)_plus}) and (\ref{Eqn:xi(n+1)_plus_in_terms_of_xi(0)_plus}) followed by imposition of requirements (\romannumeral 1),(\romannumeral 2) and (\romannumeral 3) yields the following expressions
\begin{widetext}
\begin{align}
    \tilde{Q}^{+}_{i}(0)=\frac{\alpha^{2}\Big[\sum\limits_{l=0}^{p+q-1}\Theta(l,p+q-1)C_{s(l),i}e^{-\alpha\tilde{\Delta}(l,p+q-1)}+C_{s(p+q),i}\Big]}{\Big(1-e^{-\alpha\tilde{\Delta}(0,p+q-1)}\Big)}\quad,
    \label{Eqn:Qi(0)tilda}
\end{align}
\begin{align}
    \tilde{E}^{+}_{i}(0)=\frac{
 \Big[\tilde{Q}^{+}_{i}(0)\tilde{\Delta}(0,p+q-1)e^{-\alpha\tilde{\Delta}(0,p+q-1)}+\alpha^{2}\sum\limits_{l=0}^{p+q-1}\Theta(l,p+q-1)C_{s(l),i}e^{-\alpha\tilde{\Delta}(l,p+q-1)}\tilde{\Delta}(l,p+q-1)\Big]}{\Big(1-e^{-\alpha\tilde{\Delta}(0,p+q-1)}\Big)}\quad,
    \label{Eqn:Ei(0)tilda}
\end{align}
\begin{align}
 \tilde{x}_{i}^{+}(0)=\frac{1}{\Big(1-P_i(0,p+q)e^{-\tilde{\Delta}(0,p+q-1)} \Big)}\Bigg[\sum\limits_{l=0}^{p+q-1}\Theta(l,p+q-1)
    \zeta_{i}(\tilde{\tau}(l-1),  \tilde{E}^{+}_{i}(l-1),\tilde{Q}^{+}_{i}(l-1))e^{-\tilde{\Delta}(l,p+q-1)}P_{i}(l,p+q)\nonumber \\ +P_{i}(p+q,p+q)\zeta_{i}(\tilde{\tau}(p+q-1),\tilde{E}^{+}_{i}(p+q-1),\tilde{Q}^{+}_{i}(p+q-1))\Bigg] \quad,
    \label{Eqn:xi(0)tilda}
\end{align}
\end{widetext}
where $\tilde{\Delta}(l,p+q-1)=\sum\limits_{j=l}^{p+q-1}\tilde{\tau}(j)$.
It is clear from the above expressions of $\tilde{Q}^{+}_{i}(0)$ and $\tilde{E}^{+}_{i}(0)$ that they can be obtained completely from $\{\tilde{\tau}(m)\}_{m=0}^{p+q-1}$. Note that $g_i$ are contained inside $\zeta_{i}$ but not explicitly indicated. $\tilde{Q}^+_{i}(m)$ and $\tilde{E}^+_{i}(m)$ for $m=1,2,\cdots,p+q-1$ can also be expressed entirely in terms of $\{\tilde{\tau}(m)\}_{m=0}^{p+q-1}$, starting from $\tilde{Q}^+_i(0)$, $\tilde{E}^+_i(0)$ in Eqs. (\ref{Eqn:Qi(n+1)_plus_in_terms_of_Qi(0)_plus}) and (\ref{Eqn:Ei(n+1)_plus_in_terms_of_Ei(0)_plus}). 
Replacing the resulting expressions into Eq.(\ref{Eqn:xi(0)tilda}) will also express $\tilde{x}_{i}^{+}(0)$ in terms of $\{\tilde{\tau}(m)\}_{m=0}^{p+q-1}$ alone.  Next, from Eq.(\ref{Eqn:xi(n+1)_plus_in_terms_of_xi(0)_plus}) we can obtain each $\tilde{x}_{i}^{+}(m)$ in terms of $\{\tilde{\tau}(m)\}_{m=0}^{p+q-1}$.
Let us define a vector $\mathbf{\tilde{X}}^{+}_{i}(m)=[\tilde{x}^{+}_{i}(m),\tilde{E}^{+}_{i}(m),\tilde{Q}^{+}_{i}(m)]$ for neuron $i$. It follows that the 
$(p+q)$ periodic values of $\mathbf{\tilde{X}}^{+}_{i}(m)$  can be expressed entirely in terms of  $(p+q)$ undetermined $\{\tilde{\tau}(m)\}_{m=0}^{p+q-1}$ and system parameters.\\
\\
\textbf{Step 3:} So far neurons in the periodic spike sequence $\{s(m)\}_{m=0}^{p+q-1}$ are forcibly reset after a time interval $\{\tilde{\tau}(m)\}_{m=0}^{p+q-1}$  irrespective of voltage values of the neuron $s(m)$. However, $\tilde{\tau}$'s cannot be arbitrary. Neuron $s(m+1)$ should be reset only at that $\tilde{\tau}(m)$ at which $\tilde{x}_{s(m+1)}^{-}(m+1)$ reaches the threshold. Hence, from Eq.(\ref{Eqn:eqnfortau_i})
\begin{align}
\nonumber
    {x}^{-}_{s(m+1)}(m+1)=\mathcal{F}_{s(m+1)}(\mathbf{\tilde{X}}^{+}_{s(m+1)}(m),\tilde{\tau}(m))=1, \\ m=0,1,\cdots,p+q-1.
    \label{Eqn:xs(n+1)minus=1}
\end{align}
 Substituting $m=0,1,2,\ldots,(p+q-1)$ in Eq.(\ref{Eqn:xs(n+1)minus=1})  for the neuron firing the $m^{th}$ spikes, as per the scheduled periodic spike sequence, we get
\begin{align}
\nonumber
    x_{s(1)}^{-}(1)&=\mathcal{F}_{s(1)}(\mathbf{\tilde{X}}^{+}_{s(1)}(0),\tilde{\tau}(0))=1\\
    \nonumber
       x_{s(2)}^{-}(2)&=\mathcal{F}_{s(2)}(\mathbf{\tilde{X}}^{+}_{s(2)}(1),\tilde{\tau}(1))=1\\
       \nonumber
       \vdots\\
       \nonumber
          x_{s(p+q)}^{-}(p+q)&=\mathcal{F}_{s(p+q)}(\mathbf{\tilde{X}}^{+}_{s(p+q)}(p+q-1),\tilde{\tau}(p+q-1))\\ \nonumber
          &\qquad \qquad \qquad= 1 \\
            \label{Eqn:tau_eqns}
\end{align}
Note that we can replace  $s(p+q)=s(0)$ and $x^-_{s(p+q)}(p+q)=x^-_{s(0)}(0)$, by periodicity, in the last equation. We know from Eqs.(\ref{Eqn:Qi(n+1)_plus_in_terms_of_Qi(0)_plus}),(\ref{Eqn:Ei(n+1)_plus_in_terms_of_Ei(0)_plus}), and (\ref{Eqn:xi(n+1)_plus_in_terms_of_xi(0)_plus}) along with Eqs.(\ref{Eqn:Qi(0)tilda})-(\ref{Eqn:xi(0)tilda}) that every $\tilde{\mathbf{X}}^+_i(m)$ for each $i$ can be expressed entirely in terms of $\{\tilde{\tau}(m)\}_{m=0}^{p+q-1}$, i.e.
$\mathbf{\tilde{X}}^+_i(m)=\mathcal{G}_i^m(\tilde{\tau}(0),\tilde{\tau}(1),\cdots,\tilde{\tau}(p+q-1))$. Substituting these into Eq.(\ref{Eqn:tau_eqns}) results in $(p+q)$ equations in the $(p+q)$ unknowns $\tilde{\tau}(0),\tilde{\tau}(1),\ldots,\tilde{\tau}(p+q-1)$: 
\begin{eqnarray}
\nonumber
    \mathcal{H}_{0}(\tilde{\tau}(0),\tilde{\tau}(1),\cdots,\tilde{\tau}(p+q-1))=0\\
\nonumber
    \mathcal{H}_{1}(\tilde{\tau}(0),\tilde{\tau}(1),\cdots,\tilde{\tau}(p+q-1))=0\\
\nonumber
\vdots\\
\label{Eqn:tau_eqns_H(tau)}
 \mathcal{H}_{p+q-1}(\tilde{\tau}(0),\tilde{\tau}(1),\cdots,\tilde{\tau}(p+q-1))=0
\end{eqnarray}
We numerically solve these equations for $\tilde{\tau}(m)'s$. The resulting $\{\tilde{\tau}(m)\}_{m=0}^{p+q-1}$ will in turn completely determine the $\{\mathbf{\tilde{X}}_{i}(m)\}_{m=0}^{p+q-1}$ with $i=1,2$. There may exist unphysical $\{\tilde{\tau}(m)\}_{m=0}^{p+q-1}$ and $\{\tilde{\mathbf{X}}_i(m)\}_{m=0}^{p+q-1}$ solutions as we explain next.
 \\\\
  \textbf{Step 4:} Firstly, extended voltage trajectory of neuron $1$ (which receives an inhibitory current) can cross the threshold at multiple points (see Sec.\ref{sec:Eqn_of_motion}). Each crossing occurs at times given by the solutions of $\mathcal{F}_1(\mathbf{\tilde{X}}^{+}_{1}(m),u)=1$ for $u$, with the first crossing at $\tau_1^*(m)$. $\tilde{\tau}(m)$ obtained from Eq.(\ref{Eqn:tau_eqns_H(tau)})
 will be equal to one of these solutions, if $s(m+1)=1$. It could turn out to be greater than $\tau_{1}^*(m)$ (see Fig.\ref{fig:Conditions1and2}). Since the actual trajectory will reset at $\tau^*_1(m)$, a solution $\tilde{\tau}(m)\neq \tau_1^*(m)$ will be unphysical. The procedure, so far, does not remove these unphysical solutions. The voltage trajectory of neuron $2$ (which receives an excitatory current), on the other hand, crosses the threshold only once, ensuring that $\tilde{\tau}(m)=\tau_{2}^*(m)$ if $s(m+1)=2$. It follows from above that the following condition must be imposed on solutions of Eq.(\ref{Eqn:tau_eqns_H(tau)}) at every $m$.
 \\\\
\textbf{Condition 1:} $\tilde{\tau}(m)=\tau^*_{s(m+1)}(m)$.
\\ \\
 Secondly, let us identify the periodic sequence of neurons that did not fire the $m^{th}$ spike by $\{r(m)\}_{m=0}^{p+q-1}$. For example, if $\{s(m)\}_{m=0}^{4}\{1,2,2,1,2\}$ then $\{r(m)\}_{m=0}^{4}=\{2,1,1,2,1\}$. Let us denote the first crossing of the threshold by neuron $r(m+1)$ as $\tau^*_{r(m+1)}(m)$. 
Once the solutions $\tilde{\mathbf{X}}_i^{+}(m)$ are known at the end of step $3$, $\tau^*_{r(m+1)}(m)$ will be found as the smallest positive root of $\mathcal{F}_{r(m+1)}(\tilde{\mathbf{X}}^+_{r(m+1)}(m),\tau)=1$. To ensure that $\{s(m)\}_{m=0}^{p+q-1}$ forms a valid periodic spike sequence $r(m)$'s first threshold crossing should occur later than the threshold crossing of $s(m)$ identified as the solution in step $3$. Therefore, the solutions of Eq.(\ref{Eqn:tau_eqns_H(tau)}) for $\tilde{\tau}(m)$ must satisfy the following conditions at every $m$:
\\\\
\textbf{Condition 2:} 
$\tau^*_{r(m+1)}(m)>\tilde{\tau}(m)$\\\\
Fig.\ref{fig:Conditions1and2} shows an illustrative case where condition 2 is not met. If either of the above two conditions is not met by the solutions of Eqs.(\ref{Eqn:tau_eqns_H(tau)}), it should be rejected. In that case, a $p/q$ solution with the \textit{specified periodic spike sequence} will not  exist at the corresponding set of parameters $(g,\alpha)$. In practice, the $\tau^*$'s in the above  conditions need to be found numerically by locating the first positive root of Eq.(\ref{Eqn:eqnfortau_i}). In Appendix \ref{appendix:Firing_Death} we illustrate the above formalism by applying it to  the case of $0/1$ frequency-locking in an E-I pair, where explicit solutions of Eqs.(\ref{Eqn:Qi(0)tilda})-(\ref{Eqn:xi(0)tilda}),(\ref{Eqn:tau_eqns_H(tau)}) are found.
\par 
Earlier approaches  \cite{Coombes1999,van_vreeswijk_1994,van1996partial} for determining frequency locked states had set up equations for firing times of individual neurons. These equations were solved for identical E-E and I-I pair in \cite{van_vreeswijk_1994} yielding in-phase, anti-phase, and out of phase states. However, for the case of I-I pairs, since inhibitory currents are involved, equations for firing times can have multiple solutions leading to violation of condition $1$ or $2$. In order to determine whether either of these conditions are violated equation for $\tilde{\mathbf{X}}_i$ needed to be part of the formalism. The procedure in \cite{van_vreeswijk_1994} is extended in \cite{Coombes1999} for high-order frequency locking. In addition to firing-time equations, in this case likewise, an additional $p+q$ equations for $\tilde{\mathbf{X}}_i$'s should be constructed in order to determine whether the above two conditions are satisfied. In our approach, these equations are an integral part of the formalism.

\subsection{Linear stability Analysis}
\label{subsec:Linear_Stability}
In this section we develop a procedure to find the linear stability of a valid periodic spike sequence $\{s(m)\}_{m=0}^{p+q-1}$ for a $p/q$ frequency-locked state. In \cite{van_vreeswijk_1994,Bressloff1998} stability analysis for $1/1$ state was carried out by constructing firing time perturbation equations. In  \cite{Coombes1999} this approach was extended to high-order frequency locking and resulted in an infinite dimensional eigenvalue problem. The problem was rendered tractable for \textit{non-identical} neurons under \textit{weak} coupling. In this section we develop a method, based on event-driven maps, without these restrictions. This method will be able to tackle the case of \textit{identical} neurons which is the focus of our work. Previously, event-driven maps have been used to determine the linear stability of splay states in a network of LIF neurons \cite{Zillmer2006}. This procedure is not applicable for $p/q$ frequency locking, where the spiking order of the two neurons is not preserved.

\par
Let us begin by expressing the right hand side of Eqs.(\ref{Eqn:x_plus_map})-(\ref{Eqn:Q_plus_map}) by  $f^x_i$ ,
$f^E_i$, and  $f^Q_i$ respectively. To prevent overloading of notation we will now denote $s(n+1)$ as merely $s$ and suppress superscript $\lq + \rq$ on all variables unless specified otherwise.
Each of $f^x_i$ ,
$f^E_i$, and $f^Q_i$ have an explicit dependence on $x_{i}(n),E_{i}(n),Q_{i}(n)$, and $\tau(n)$. They also have an implicit dependence, through $\tau(n)$, on the current values of $x_{s}(n),E_{s}(n)$, and $Q_{s}(n)$ of the neuron $s$ that will fire the $(n+1)^{th}$ spike. Following the general linear stability procedure, perturbations of Eqs.(\ref{Eqn:x_plus_map})-(\ref{Eqn:Q_plus_map}) will evolve as
\begin{align}
\nonumber
    \delta x_{i}(n+1)=&\frac{\partial f_{i}^x}{\partial x_{i}}\delta x_{i}(n)+\frac{\partial f_{i}^x}{\partial E_{i}}\delta E_{i}(n)+\frac{\partial f_{i}^x}{\partial Q_{i}}\delta Q_{i}(n)\\
    &+\frac{\partial f_{i}^x}{\partial \tau}\delta \tau(n) \quad,
        \label{Eqn:x_perturbation_1}
\end{align}
\begin{align}
    \delta E_{i}(n+1)=\frac{\partial f^E_i}{ \partial E_i}\delta E_i(n)+\frac{\partial f^E_i}{ \partial Q_i}\delta Q_i(n)+\frac{\partial f^E_i}{\partial \tau}\delta \tau(n) \quad,
\label{Eqn:E_perturbation}
\end{align}

\begin{align}
\delta Q_i(n+1)=\frac{\partial f^Q_i}{\partial Q_i}\delta Q_i(n)+\frac{\partial f^Q_i}{\partial \tau}\delta \tau(n) \quad ,
\label{Eqn:Q_perturbation_1}
\end{align}
where the partial derivatives are evaluated at values held by the state variables soon after the $n^{th}$ spike.
Since the identity of neuron $s$ is known, $\tau(n)$ can be obtained from Eq.(\ref{Eqn:eqnfortau_i})
by setting $\mathcal{F}_{s}(x_{s}(n),E_{s}(n),Q_{s}(n),\tau(n))=1$. The perturbation $\delta\tau(n)$ can be obtained by writing the differential of $\mathcal{F}_s$ and solving for $\delta\tau(n)$ yielding
\begin{align}
\nonumber
    \delta \tau(n)=-\Bigg({\frac{\partial \mathcal F_s}{\partial \tau}}\Bigg)^{-1}\Bigg( \frac{\partial \mathcal F_s}{\partial x_s}\delta x_{s}(n)+ \frac{\partial \mathcal F_s}{\partial E_s}\delta E_{s}(n)\\+ \frac{\partial \mathcal F_s}{\partial Q_s}\delta Q_{s}(n)\Bigg) .
\end{align}
Clearly $\delta \tau$ depends only on the state of neuron $s$. Replacing for $\delta \tau(n)$ in Eqs.(\ref{Eqn:x_perturbation_1})-(\ref{Eqn:Q_perturbation_1}) 
\begin{align}
    \delta \mathbf{X}(n+1)=\Big(\mathbf{DF}(n) \Big)_{s(n+1)} \delta \mathbf{X}(n) ,
    \label{Eqn:delta X(n+1)}
\end{align}
where  $\mathbf{X}=[\mathbf{X}_1, \mathbf{X}_2]^{\mathbf{T}}$ represents the complete state vector after the $n^{th}$ spike and $\mathbf{F}=[\mathbf{F}_1,\mathbf{F}_2]^{\mathbf{T}}$ where $\mathbf{F}_i=[f^x_i,f^E_i,f^Q_i]$. For $s(n+1)=1$, the Jacobian matrix has a block lower triangular form
\begin{equation}
\Big(\mathbf{DF}(n)\Big)_1=
    \begin{bmatrix}
    \mathbf{R_1} & \mathbf{0}\\
    \mathbf{B_1} & \mathbf{U_1}\\ 
\end{bmatrix} ,
\end{equation}
and for $s(n+1)=2$, it takes the form of a block upper triangular form
\begin{equation}
\Big(\mathbf{DF}(n)\Big)_2=
    \begin{bmatrix}
    \mathbf{U_2} & \mathbf{B_2}\\
    \mathbf{0} & \mathbf{R_2}\\
\end{bmatrix} ,
\end{equation}
where $\mathbf{0}$ is $3\times 3$ null matrix with,
\begin{align}
\mathbf{R_1}=
\begin{bmatrix}
0 & 0 & 0 \\[1em]
\overline{\mathcal{F}}_{1,x_1}\frac{\partial f^E_{1}}{\partial \tau}  &  \frac{\partial f^E_{1}}{\partial E_1} + \overline{\mathcal{F}}_{1,E_1}\frac{\partial f^E_{1}}{\partial \tau} & \frac{\partial f^E_{1}}{\partial Q_1} + \overline{\mathcal{F}}_{1,Q_1}\frac{\partial f^E_{1}}{\partial \tau} \\[1em]
\overline{\mathcal{F}}_{1,x_1}\frac{\partial f^Q_{1}}{\partial \tau} & \overline{\mathcal{F}}_{1,E_1}\frac{\partial f^Q_{1}}{\partial \tau} & \frac{\partial f^Q_{1}}{\partial Q_1} + \overline{\mathcal{F}}_{1,Q_1}\frac{\partial f^Q_{1}}{\partial \tau}\\[0.5em]
\end{bmatrix} \quad,
\label{Eqn:R1matrix}
\end{align}
\begin{align}
\mathbf{B_1}=
 \begin{bmatrix}
\overline{\mathcal{F}}_{1,x_1}{\partial f^x_{2} \over \partial \tau}  &\quad  \overline{\mathcal{F}}_{1,E_1}{\partial f^x_{2} \over \partial \tau}  & \quad  \overline{\mathcal{F}}_{1,Q_1}{\partial f^x_{2} \over \partial \tau} \quad  \\[1em]
\overline{\mathcal{F}}_{1,x_1}{\partial f^E_{2} \over \partial \tau} \ & \quad \overline{\mathcal{F}}_{1,E_1}{\partial f^E_{2} \over \partial \tau}  & \quad  \overline{\mathcal{F}}_{1,Q_1}{\partial f^E_{2} \over \partial \tau} \quad  \\[1em]
\overline{\mathcal{F}}_{1,x_1}{\partial f^Q_{2} \over \partial \tau}  \quad  & \quad
\overline{\mathcal{F}}_{1,E_1}{\partial f^Q_{2} \over \partial \tau}   & \quad \overline{\mathcal{F}}_{1,Q_1}{\partial f^Q_{2} \over \partial \tau} \\[0.5em]
\end{bmatrix} 
\label{Eqn:B1_matrix} ,
\end{align}
and
\begin{align}
\mathbf{U_1}=
\begin{bmatrix}
{\partial f^x_{2} \over \partial x_2} \quad &\quad {\partial f^x_{2} \over \partial E_2} \quad &\quad {\partial f^x_{2} \over \partial Q_2} \\[1em]
0 \quad & \quad {\partial f^E_{2} \over \partial E_2}\quad &\quad  {\partial f^E_{2} \over \partial Q_2}\\[1em]
0 \quad&\quad 0 \quad&\quad {\partial f^Q_{2} \over \partial Q_2}\\[1em]
\end{bmatrix} ,
\label{Eqn:U1_matrix}
\end{align} 
where 
$\overline{\mathcal{F}}_{1,x_1}\mspace{-5mu}=\mspace{-5mu}-\left( \frac{\partial \mathcal{F}_1}{\partial x_1} \right) \left( \frac{\partial \mathcal{F}_1}{\partial \tau} \right)^{-1}$, $\overline{\mathcal{F}}_{1,E_1}\mspace{-5mu}=\mspace{-5mu} -\left( \frac{\partial \mathcal{F}_1}{\partial E_1} \right) \left( \frac{\partial \mathcal{F}_1}{\partial \tau} \right)^{-1}$, and
$\overline{\mathcal{F}}_{1,Q_1} = -\left( \frac{\partial \mathcal{F}_1}{\partial Q_1} \right) \left( \frac{\partial \mathcal{F}_1}{\partial \tau} \right)^{-1}$. The corresponding $\mathbf{R_2},\mathbf{B_2}$ and $\mathbf{U_2}$ will be obtained by replacing the indices $1$ with $2$ and vice-versa in Eqs.(\ref{Eqn:R1matrix})-(\ref{Eqn:U1_matrix}).
To examine the evolution of perturbations in a $p/q$ frequency-locked state, extending Eqs.(\ref{Eqn:delta X(n+1)}), to connect perturbations separated by $(p+q)$ spikes yields
 \begin{widetext}
\begin{align}
    \delta \tilde{\mathbf{X}}(p+q)=\Big(\mathbf{DF}(p+q-1) \Big)_{s(p+q)}\Big(\mathbf{DF}(p+q-2) \Big)_{s(p+q-1)}\cdots\Big(\mathbf{DF}(0) \Big)_{s(1)} \delta \tilde{\mathbf{X}}(0) .
     \label{Eqn:delta_X(p+q)}
\end{align}
\label{Eqn:product_of_jacobians}
\end{widetext}
The perturbations are connected by a product of $(p+q)$ Jacobians. The Jacobian $\Big(\mathbf{DF}(m)\Big)_{s(m+1)}$ is calculated at $\tilde{\mathbf{X}}(m)$
. The stability of $p/q$ frequency locking is found by computing the eigenvalues of the matrix formed from the product of Jacobians in the above equation. If absolute value of at least one eigenvalue is greater than unity then $p/q$ frequency-locked state is unstable. Otherwise it is stable.
\par

\subsection{Non-smooth Bifurcations}
\label{subsec:nonsmooth_bifurcations}
Non-smooth bifurcations in discontinuous systems have an extensive literature \cite{PWS2008}. Such bifurcations can lead to changes which cannot occur through smooth bifurcations. For instance, period $1$ can go to period $3$ and then to chaos in a discontinuous, piecewise linear map \cite{JAIN2003}. In another example, chaos can emerge from a fixed point attractor \cite{Nusse1992} in a continuous, piecewise smooth map. 
One class of nonsmooth bifurcations, known as grazing bifurcations, are observed in threshold systems and can lead to the disappearance of a single solution without any other changes. Forced integrate-and-fire neurons are known to display particular types of grazing bifurcations \cite{Coombes2001,Coombes2001_errata}. At a grazing bifurcation, the underlying voltage trajectory of a neuron is tangent to the threshold \cite{Coombes2001,KhajehAlijani2009}.  This tangency is sometimes referred to as a graze in literature.  
\begin{figure*}[t]
    \centering
    \includegraphics[scale=0.3]{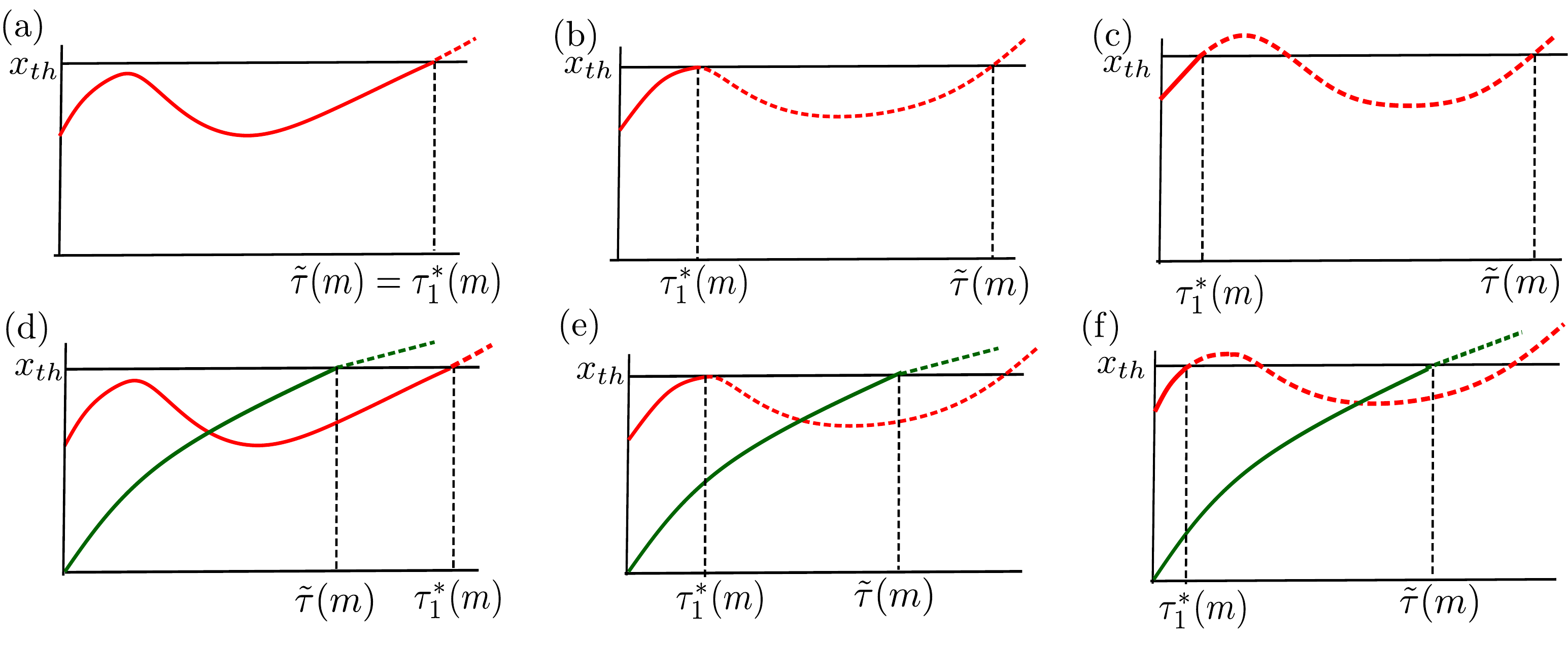}
    \caption{(Color online) Schematic figures displaying grazing bifurcation of Type-1 in (a)-(c) and Type-3 in (d)-(f) on varying a parameter. Extended voltage trajectories of excitatory and inhibitory neurons are shown in red and green respectively. In (a)-(c) the trajectory of an inhibitory neuron is not shown, as it does not play a role. Solid lines represent physical portions of the trajectory whereas dashed lines represent unphysical portions.}
    \label{fig:type1_and3_bif}
\end{figure*}
\begin{figure*}[t]
    \centering
    \includegraphics[scale=0.2]{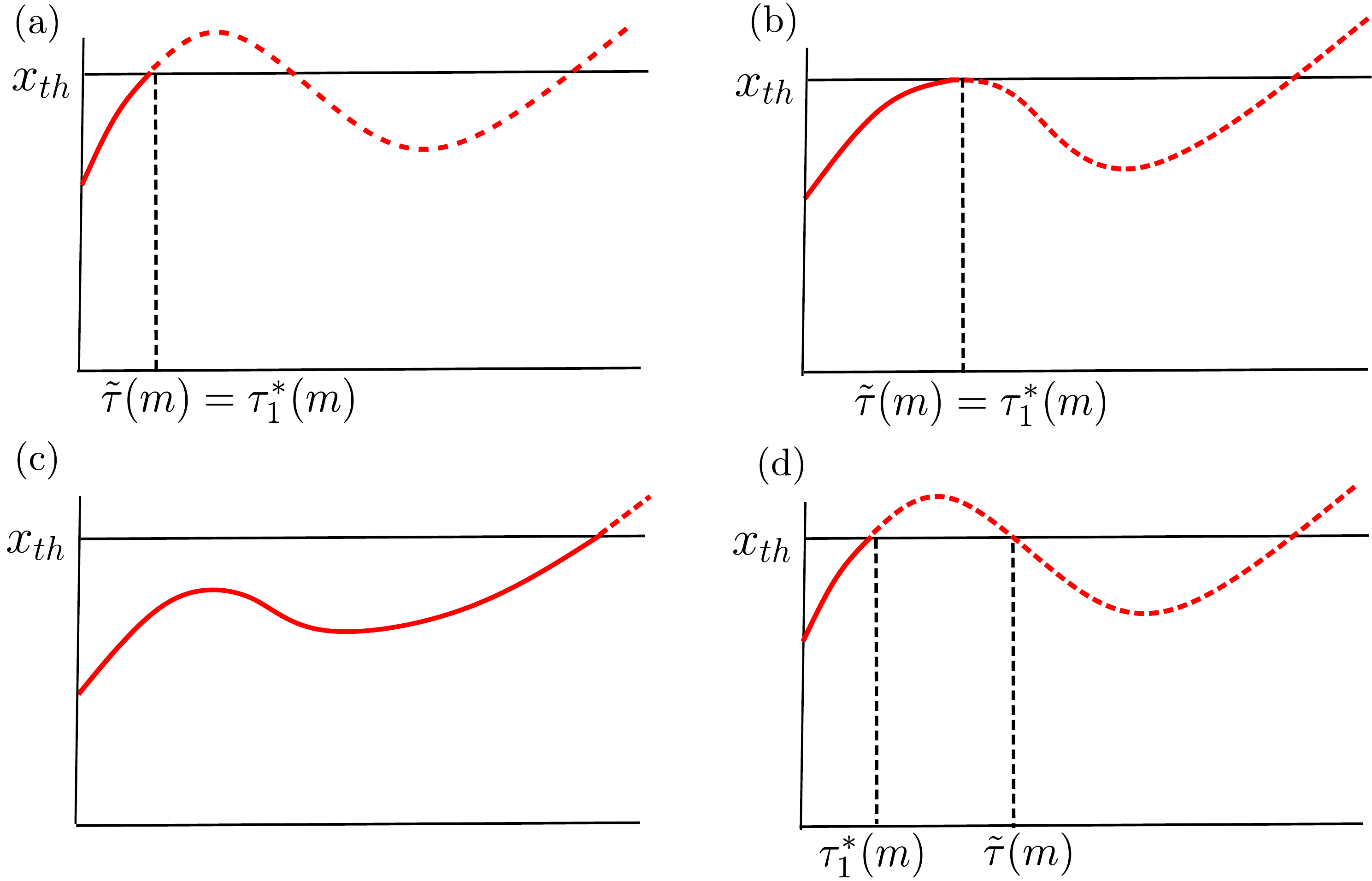}
    \caption{(Color online) Schematic figures of extended voltage trajectories of excitatory neuron displaying grazing bifurcations of Type-2a and Type-2b on varying a parameter. In Type-2a bifurcation (a),(b) is followed by (c).  In Type-2b bifurcation (a),(b) is followed by (d). Inhibitory neuron trajectory does not play a role, thus not shown.}
    \label{fig:type2_bif}
\end{figure*}
\par
In an E-I pair, the inhibitory neuron, which receives an excitatory current, cannot graze the threshold since its voltage increases monotonically. A grazing bifurcation will occur when the voltage trajectory of the excitatory neuron is tangent to the threshold. The graze can occur between  any two consecutive spikes, $m^{th}$ and $(m+1)^{th}$, in the periodic spike sequence $\{s(m)\}_{m=0}^{p+q-1}$. To describe grazing bifurcations of different types, we must consider an extended voltage trajectory obtained by ignoring the reset condition after the $m^{th}$ spike. We need to consider two distinct cases that result in different sets of grazing bifurcations. In case A, both the $(m+1)^{th}$ spike and the graze arises from the excitatory neuron. In case B, $(m+1)^{th}$ spike is due to inhibitory neuron, i.e., $s(m+1)=2$, whereas the graze originates from the excitatory neuron. 
\par
In order to uncover various mechanisms of grazing bifurcation for case A, we need to further consider two separate situations. In situation (\romannumeral1), on varying a parameter, the local maxima $x_{max}$ of the excitatory neuron's extended voltage trajectory rises towards the threshold from below, becomes tangent to it at  
some $\tau^*_{1}(m)<\tilde{\tau}(m)$, and eventually crosses it [See Fig.\ref{fig:type1_and3_bif}].  Condition $1$ of Sec.\ref{sec:pbyq_freq_locked_analytical} gets violated at the bifurcation point $(g_c,\alpha_c)$ where the graze occurs, destroying the $p/q$ solution. Thus, at the bifurcation point, for the excitatory neuron:
$\frac{d x^-_1}{d u}\Bigr\rvert_{\tilde{\tau}(m)}\neq0$ while $\frac{dx^-_1}{du}\Bigr\rvert_{\tau^*_{1}(m)}=0$ and $\scalebox{1.25}{$\sfrac{\partial x_{max}}{\partial c_k}$} \neq 0$ where $c_k=\{g,\alpha\}$. In \cite{Coombes2001} this bifurcation is termed as first type grazing bifurcation and has been earlier observed in various models of forced neurons \cite{Coombes2001_errata,KhajehAlijani2009,Coombes2012}. We will refer to it as Type-1 grazing bifurcation. 
In situation (\romannumeral2), depicted in Fig.\ref{fig:type2_bif}, parameter variation results in lowering of the local maxima from above the threshold and eventually becoming tangent to it at $\tilde\tau(m)$. Herein, $\tilde{\tau}(m)$ is located at $\tau_1^*(m)$.  Clearly, at this bifurcation point, $\frac{dx^-_1}{d u}\Bigr\rvert_{\tilde{\tau}(m)}=0$ for the excitatory neuron. On further variation of parameter, two different scenarios can come into play. In the $1^{st}$ scenario, $x_{max}$ \textit{crosses} the threshold, resulting in a loss of the solution as shown in Figs. \ref{fig:type2_bif}(a)-(c).  In this scenario, at the bifurcation point, in addition  $\scalebox{1.25}{$\sfrac{\partial x_{max}}{\partial c_k}$} \neq 0$. This bifurcation has also been observed in studies of forced neurons and is referred to as Type $2$ in \cite{Coombes2001,KhajehAlijani2009}. We will term it as Type-2a grazing bifurcation. In the $2^{nd}$ scenario, $x_{max}$ rebounds from the threshold, as illustrated through sequence of  Figs. \ref{fig:type2_bif}(a), \ref{fig:type2_bif}(b), and \ref{fig:type2_bif}(d). In this process, $\tilde{\tau}(m)$ changes its position from the first crossing, $\tau^*_{1}(m)$, to the second, violating condition ($1$). Thus, the existing $p/q$ solution is lost. In this scenario, additionally, at the bifurcation point $ \scalebox{1.25}{$\sfrac{\partial x_{max}}{\partial c_k}$} = 0$, unlike the $1^{st}$ scenario. We term this bifurcation as Type-2b grazing bifurcation. To the best of our knowledge this bifurcation is not reported elsewhere.
\par
In case B, a grazing bifurcation will occur when $x_{max}$  approaches the threshold from below, and eventually crosses it. At tangency,  $\tau^*_{1}(m)<\tilde{\tau}(m)$, meaning that excitatory neuron reaches the threshold before the inhibitory neuron, violating Condition $2$ of Sec.\ref{sec:pbyq_freq_locked_analytical}, leading to a loss of an existing $p/q$ frequency-locked solution. Clearly, at the bifurcation point $\frac{dx_1^-}{du}\Bigr\rvert_{\tau^*_{1}(m)}=0$ and $\frac{\partial x_{max}}{\partial c_k}\neq 0$ for excitatory neuron and $\frac{dx_2^-}{du}\Bigr\rvert_{\tilde{\tau}(m)}\neq0$  for inhibitory neuron.  We will designate this as Type-$3$ grazing bifurcation. This bifurcation inherently involves coupled neurons and has not been reported in any of the previous studies.
\par
Certain conclusions can be immediately drawn
about eigenvalues of the matrix connecting perturbations $(p+q)$ iterations apart (see Eq.(\ref{Eqn:delta_X(p+q)})) at Type-$2$ grazing bifurcations. Few elements in this matrix contain the derivative $
\frac{\partial \mathcal F_1}{\partial \tau}$
in the denominator evaluated at the spike time, where $\tau$ corresponds to the spike time. At the threshold, $\frac{dx_1^-}{d u}\Bigr\rvert_{u=\tau(n)}=\frac{\partial \mathcal{F}_1}{\partial \tau}\Bigr\rvert_{\tau=\tau(n)}.$ Since, $\frac{dx_1^-}{du}\Bigr\rvert_{\tilde{\tau}(m)}=0$  at a Type-2 bifurcation, the non-zero elements of $\mathbf{R_1}$ and $\mathbf{B_1}$ in Eqs.(\ref{Eqn:R1matrix})-(\ref{Eqn:B1_matrix}) blow up. The Jacobian formed from $\mathbf{R_1}$ and $\mathbf{B_1}$
is present in the product of Jacobians in Eq.(\ref{Eqn:delta_X(p+q)}). Hence, the trace of the resulting product matrix will blow up. It follows that atleast one eigenvalue of the resulting matrix will diverge at the bifurcation point. Consequently, only an \textit{unstable} frequency-locked solution can get lost through Type-$2$ bifurcations. However, Type-$1$ and Type-$3$ bifurcations can remove both stable and unstable solutions.
\par
\section{Bifurcations and multistability}
\label{sec:Bifurcations_And_multistability}
In this section, we report the results of applying the analytical method developed in Sec.\ref{sec:pbyq_freq_locked_analytical} to determine the existence, stability and bifurcations of various $p/q$ frequency-locked states.
\subsection{Firing Death ($p/q=0/1$)}
\label{sec:firing_death}
\begin{figure}[h!]
    \centering
    \includegraphics[width=0.4\textwidth]{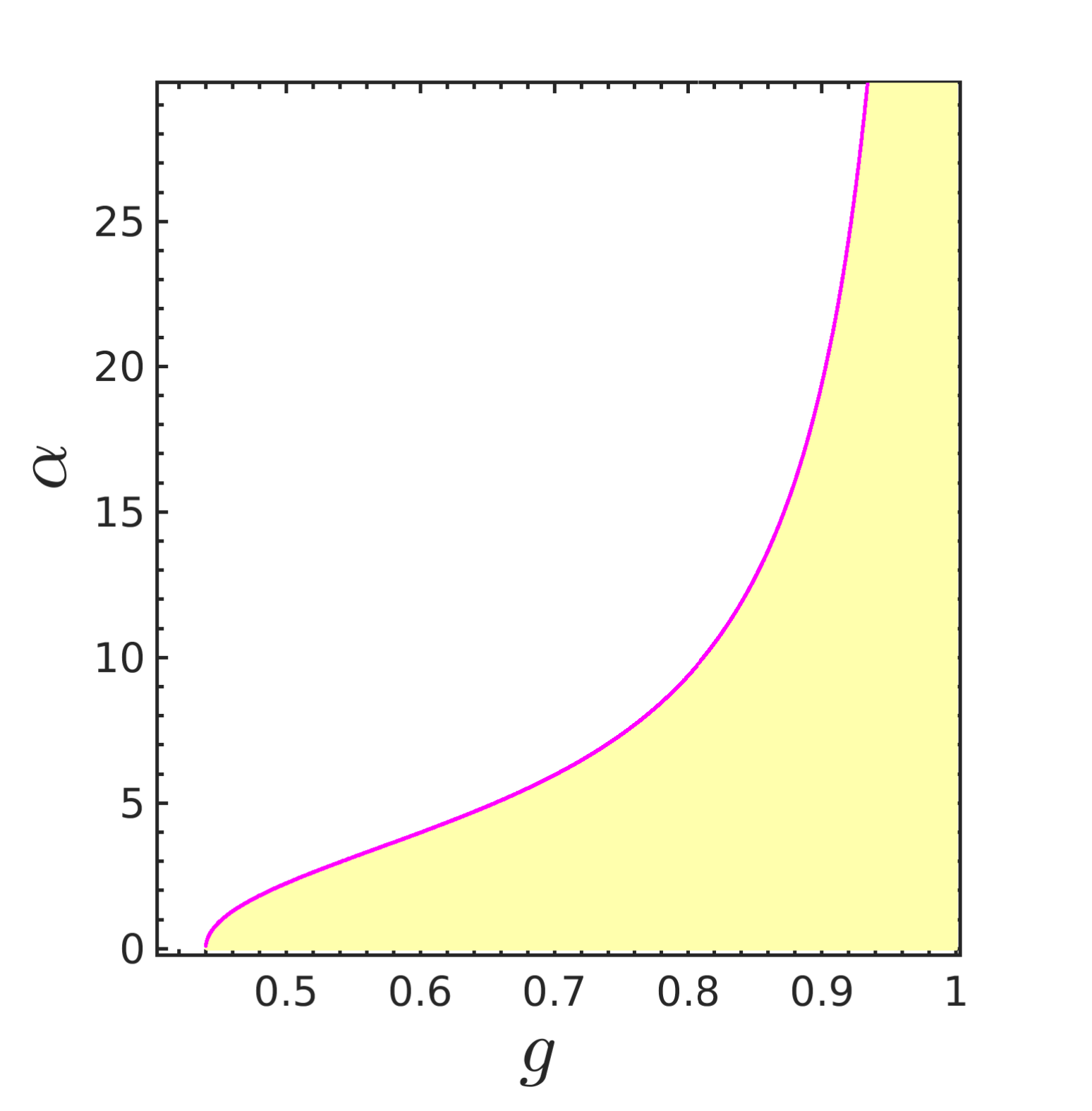}
    \caption{(Color online) Bifurcation boundary of the only spike sequence of $0/1$ firing death solution. The external boundary (magenta color) is defined by Type-3  grazing bifurcation. The yellow color belongs to the numerically obtained $0/1$ firing death solution.}
    \label{fig:FD_bif_boundary}
\end{figure}

In Appendix \ref{appendix:Firing_Death}, we illustrate the analytical method through an analysis of firing death, $0/1$ . In this case the only potential spike sequence is $\{2\}$. The equations governing frequency-locked states, for this case, undergo significant simplification,  as shown in the Appendix \ref{appendix:Firing_Death}. By solving Eq.(\ref{Eqn:tau_eqns_H(tau)}) for the lone spike sequence of $0/1$ under conditions $(1)$ and $(2)$, we obtain a \textit{single} solution. A linear stability analysis of $\{2\}$ produces a Jacobian matrix in the form of an upper block triangular matrix, whose eigenvalues can be easily obtained. All eigenvalues are found to be less than one, indicating that $0/1$ is stable across its entire region of existence. Since only a \textit{single} $0/1$ solution exists, it can be lost only through a non-smooth bifurcation.
In the appendix, we show that this solution disappears by violating condition ($2$), which implies a Type-$3$ grazing bifurcation. Additionally, we derive an equation for the curve on which this bifurcation occurs. This result is shown in Fig.\ref{fig:FD_bif_boundary} and compared with the numerically obtained boundary, demonstrating excellent agreement between the two. 
\subsection{Irreducible $p/q$ tongues with single spike sequence}
\label{subsec:tongues_with_Single_spk}
Now we describe the bifurcation structure of $p/q$ tongues which possess a single valid spike sequence. Straightforward examples of these are frequency lockings of the form $1/q$ and $q/(q+1)$ as described in Sec.\ref{sec:Eqn_of_motion}. Other $p/q$ could potentially possess multiple spike sequences arising through permutations of excitatory-inhibitory spikes. However, all permutations may not yield a physically valid periodic spike sequence. Using our analytical procedure we identified few tongues with single and multiple valid spike sequences. This was done by determining solutions of every potential spike sequence in a $p/q$ frequency locking. We determined all the \textit{valid} spike sequences for a few $2/q$ tongues. The number of valid spike sequences tends to rise with $q$ or remains the same. For $2/5,2/7,2/9,2/11,2/13$, and $2/15$ the number of valid spike sequences are $1,1,1,2,3$, and $4$ respectively. From numerical simulation results in Fig.\ref{fig:pspace_higher_resolution} we observe that on progressing from lower to higher $g$ values, the $2/q$ tongues arise in ascending order of $q$. It follows that the number of valid spike sequences tends to be greater for $2/q$ tongues that exist towards higher $g$ values. 
\begin{figure}[b]
\centering
\includegraphics[width=0.4\textwidth]{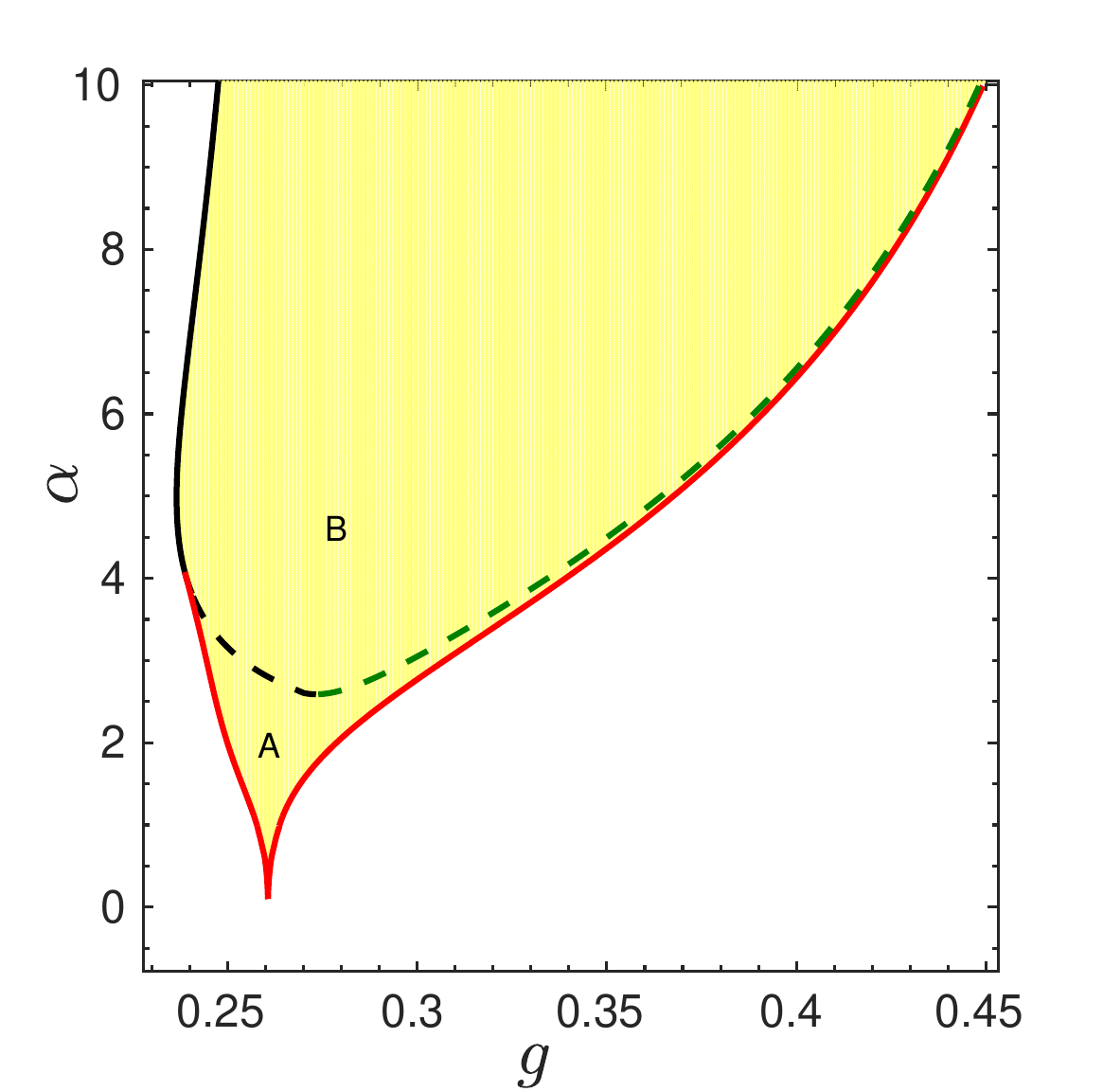}
    \caption{(Color online) Analytically obtained solutions of  $1/2$ frequency locked tongue and comparison with simulations.  Region B contains one stable solution and region A contains a pair of stable-unstable solutions of the lone spike sequence of $1/2$. Outer bifurcation boundaries of the sequence are defined by Type-1 grazing (black) and saddle-node (red) bifurcations. Inner bifurcation boundaries are defined by Type-1 and Type-2b (dark green) grazing bifurcations. Grazing bifurcation boundaries across which stable (unstable) solutions are lost are shown using solid (dashed) lines. The yellow color indicates the numerically obtained $1/2$ frequency-locked region. }
    \label{fig:bifurcation_boundaries_1by2}
\end{figure}
\par
All the tongues labeled in Fig.\ref{fig:pspace_low_Resolution_simulations} have a single spike sequence.
We will focus on $p/q=1/2$ as a representative case for all such tongues. The unique spike sequence of $1/2$ is $\{1,2,2\}$. Fig.\ref{fig:bifurcation_boundaries_1by2} shows that $1/2$ tongue is divided into regions of two solutions (labeled A) and one solution (labeled B) of $\{1,2,2\}$. The two solution region contains a  stable-unstable pair of the spike sequence. Each member of this pair has a different temporal structure. In the single solution region the spike sequence is stable. The boundary separating the regions  with different number of stable and unstable solutions of a given spike sequence will be termed as its \lq inner' boundary. On the other hand, the boundary that separates the regions of existence and absence of a spike sequence will be termed as an \lq outer' boundary of that spike sequence. In the case of $1/2$, the outer boundary of its only sequence forms the boundary of the tongue. However, as we will see in sec.\ref{subsec:tongues_with_multiple_Spk}, the relation between these boundaries is more complex for tongues that involve multiple valid spike sequences. In Fig.\ref{fig:bifurcation_boundaries_1by2}, a comparison between analytically and numerically determined $1/2$ tongue shows excellent agreement.
\par
We computed the bifurcation boundaries (both outer and inner) using continuation algorithms and found the following results depicted in Fig.\ref{fig:bifurcation_boundaries_1by2}. The outer boundaries of the sole spike sequence of $1/2$ are defined by saddle node and grazing bifurcation of Type-1. Its inner boundaries are formed by grazing bifurcations of Type-1 and Type-2b. The Type-1 grazing bifurcation on the outer boundary involves the disappearance of a single stable solution of the spike sequence. Conversely, the same bifurcation on the inner boundary involves a single unstable solution. Grazing bifurcation of  Type-2b on the inner boundary destroys the branch of unstable solution of $\{1,2,2\}$ created through a saddle node bifurcation on the $1/2$ tongue's boundary. As anticipated from the discussion in Sec. \ref{subsec:nonsmooth_bifurcations}, Type-2b grazing bifurcation entails an unstable spike sequence.
\par
\begin{figure*}
    \centering
    \includegraphics[width=0.9\textwidth]{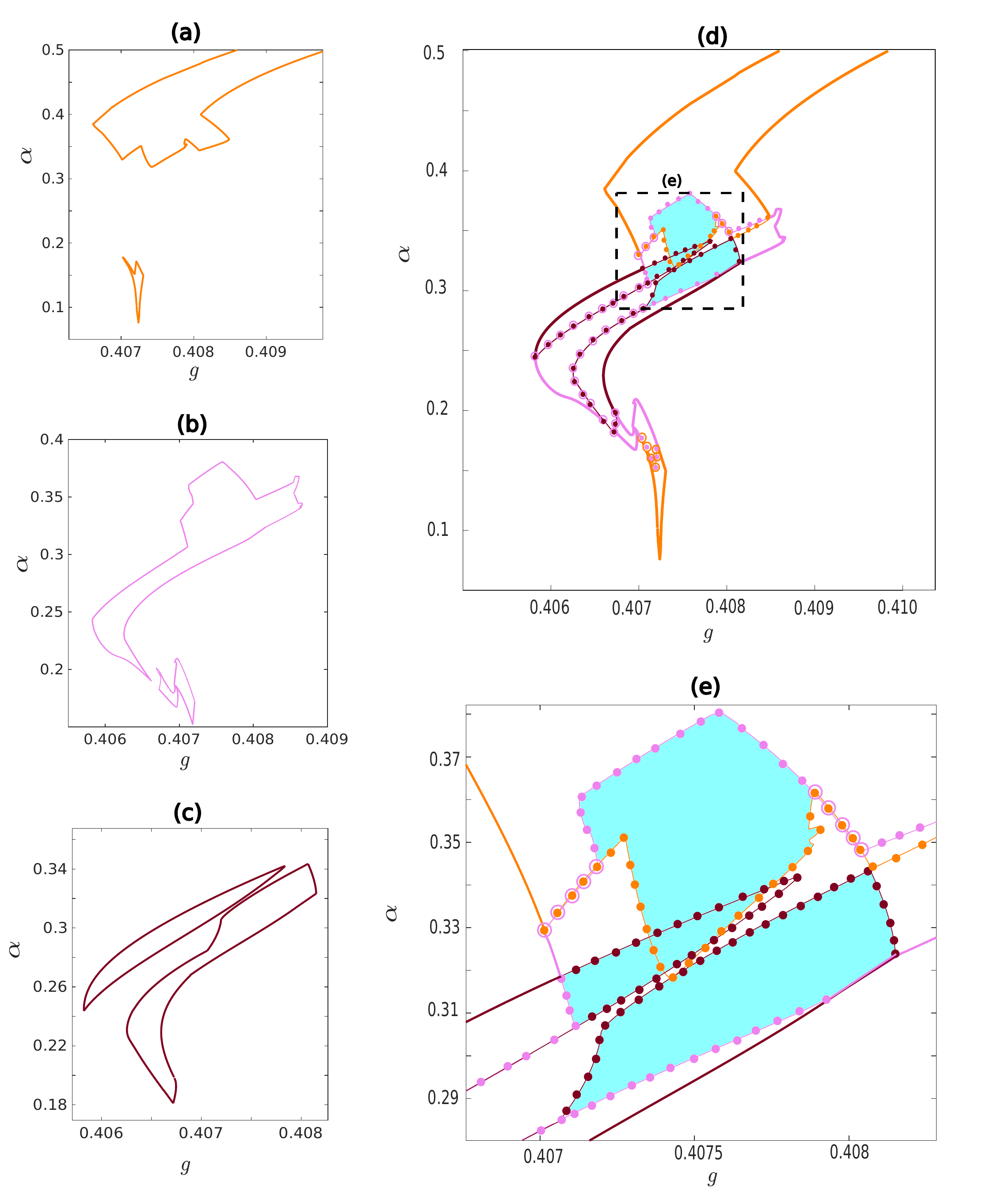}
    \caption{(Color online) The analytically obtained outer boundaries of the three valid spike sequences of $2/13$ frequency-locked state are shown in (a),(b), and (c). Spike sequence in (a),(b) and (c) are $\{1,2,^6,1,2^7\}$,  $\{1,2,^5,1,2^8\}$ and $\{1,2,^4,1,2^9\}$ respectively.  
    In (d) we combine (a),(b), and (c) to obtain the boundary of $2/13$ tongue. Panel (e) is the magnification of the boxed region in panel (d), which is labeled as (e). Boundaries of each sequence are shown by a different color. Those segments that lie on the boundary of the complete $2/13$ tongue are depicted by solid lines. Those segments that occur inside the $2/13$ region are shown (i) through dotted solid lines if no spike exchange occurs across it and (ii) through circles combined with solid dotted lines if a spike exchange occurs with the circle representing the shared boundary of one spike sequence and a dot representing that of the other. In the colored regions, regions of existence of multiple spike sequences overlap. This can lead to inter-sequence multistability.}
    \label{fig:2by13_external}
\end{figure*} 
\begin{figure*}
    \centering
    \includegraphics[width=0.9\textwidth]{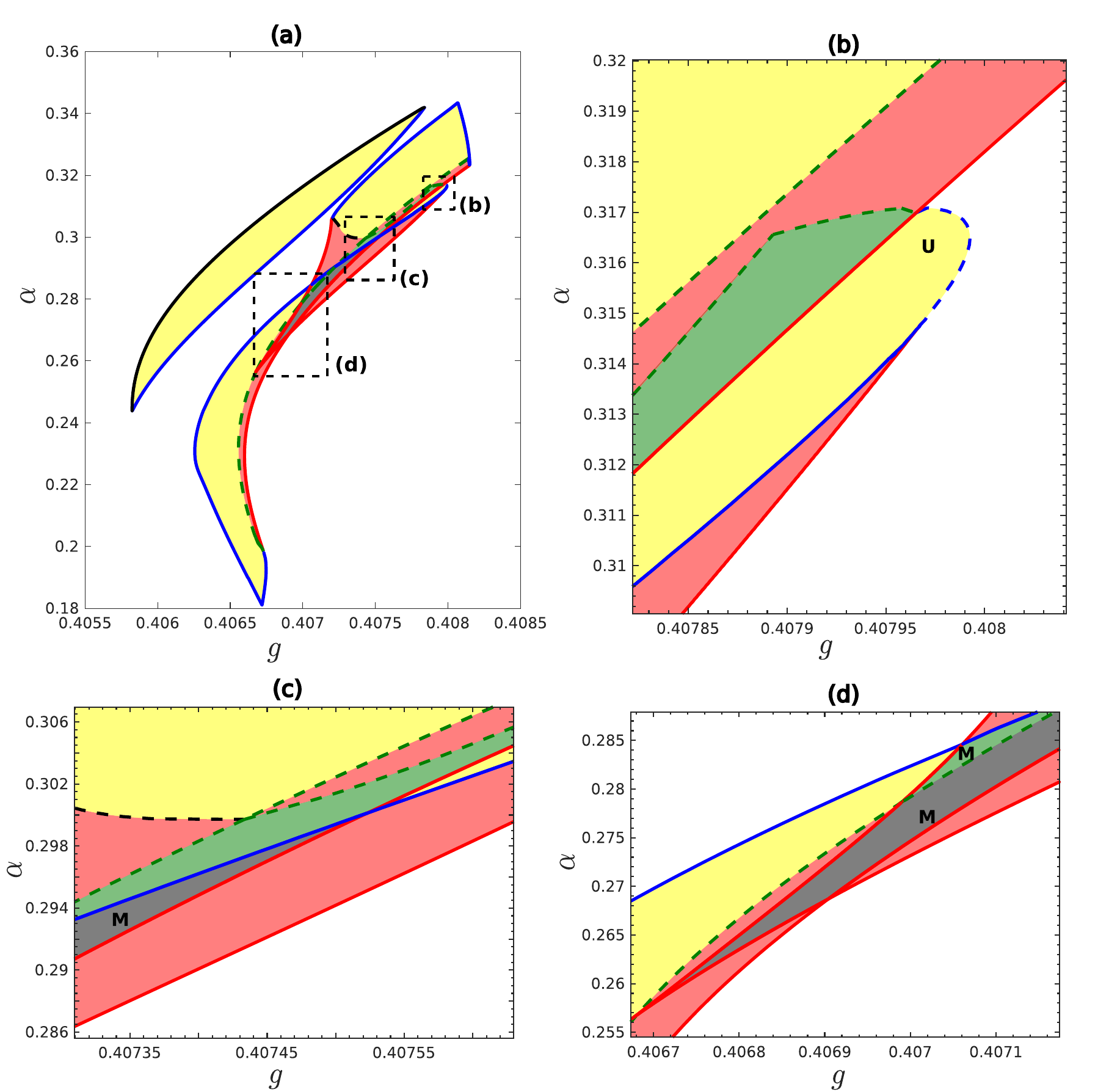}
    \caption{(Color online) This figure, analytically obtained, shows the bifurcation structure and number of solutions of the spike sequence -$\{1,2^4,1,2^9\}$- of $2/13$ frequency-locked state. Panels (b),(c), and (d) are magnifications of similarly labeled regions of panel (a). Shading colors indicate the following- yellow region: 1 solution, red region: 2 solutions, green: 3 solutions, black: 4 solution region. Regions labeled as \lq M' have intra-sequence multistability and those labeled as \lq U' have only an unstable spike sequence. The color of the curves differentiates different bifurcation types- red: saddle-node, black: Type-1 grazing bifurcation,  dark green: Type-2b grazing bifurcation. In the case of grazing bifurcations, solid (dashed) lines represent the involvement of a single stable (unstable) solution.  Blue curves are the shared exchange boundaries between two spike sequences ($\{1,2^4,1,2^9\}$ and $\{1,2^5,1,2^8\}$ in this figure).}
    \label{fig:129124_spk_sequence}
\end{figure*}
\begin{figure*}
    \centering
    \includegraphics[width=0.9\textwidth]{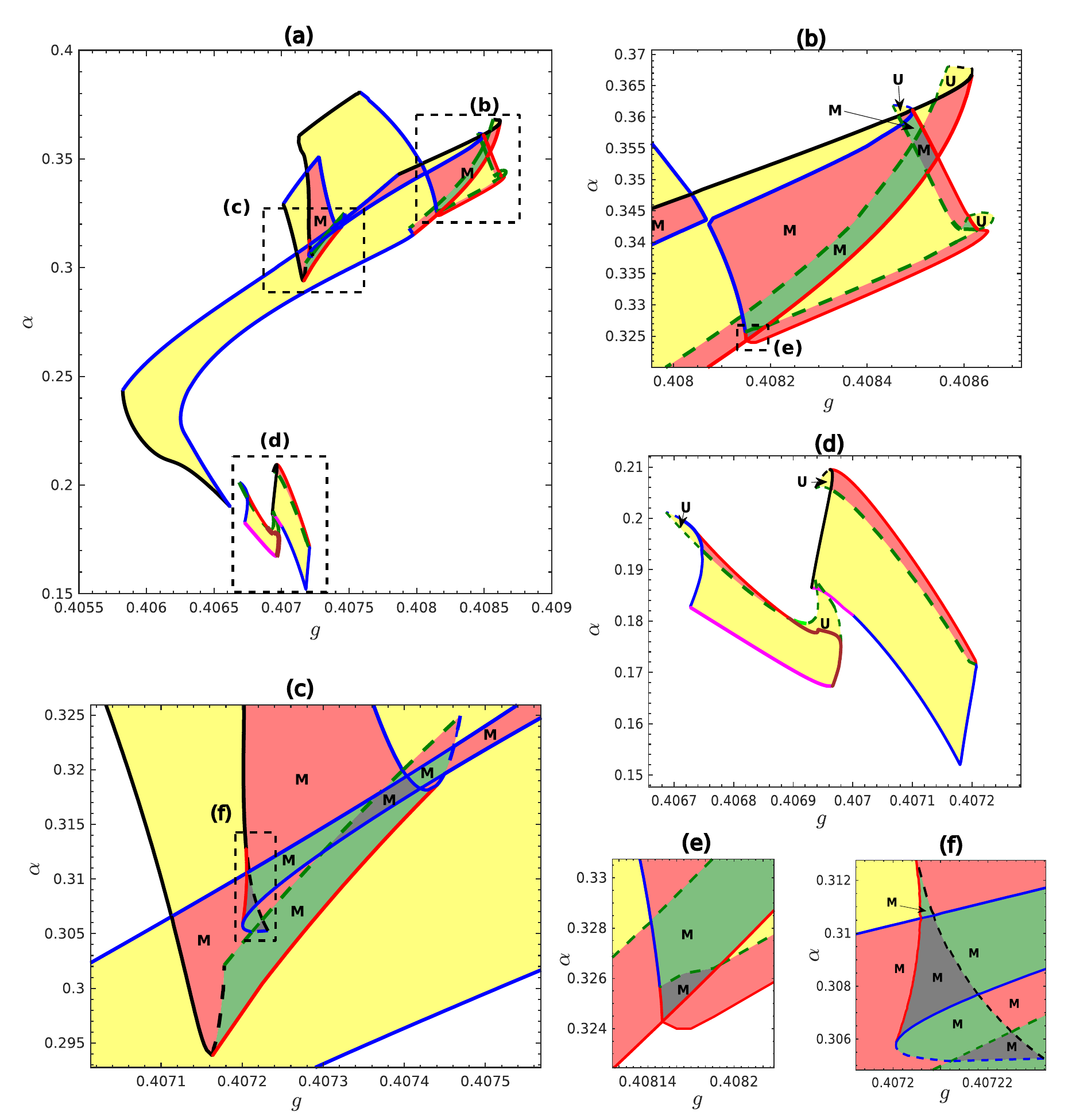}
    \caption{(Color online) This figure, analytically obtained, shows the bifurcation structure and number of solutions of the spike sequence -$\{1,2^5,1,2^8\}$- of $2/13$ frequency-locked state.  Panels (b),(c), and (d) are magnifications of similarly labeled regions of panel (a). Inset (e) and (f) are magnifications of region (e) in panel (b) and region (f) in panel (c) respectively. The color scheme, line types, and labels have the same meaning as in Fig.\ref{fig:129124_spk_sequence}. In addition, Type-3 and Type-2a grazing bifurcation are shown in magenta and light green color respectively. The brown boundary separates two regions with a single spike sequence - region \lq U' contains an unstable and other region contains a stable sequence.}   \label{fig:128125_spk_sequence}
\end{figure*}
\begin{figure*}
    \centering    \includegraphics[width=0.9\textwidth]{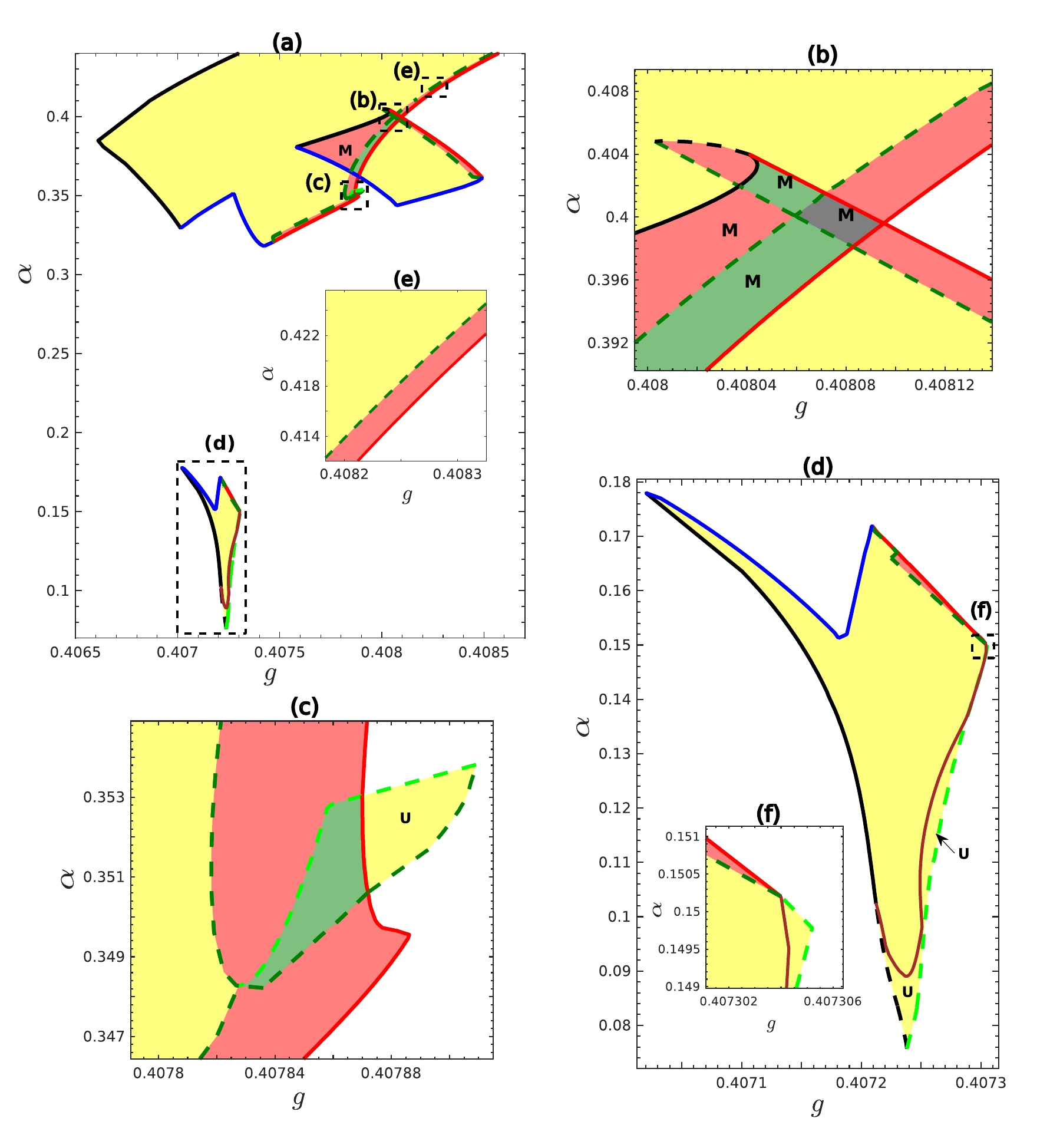}
    \caption{(Color online) This figure, analytically obtained, shows the bifurcation structure and number of solutions of the spike sequence -$\{1,2^6,1,2^7\}$- of $2/13$ frequency-locked state. Panels (b),(c), and (d) are magnifications of similarly labeled regions of panel (a). Inset (e) and (f) are magnifications of region (e) in panel (a) and region (f) in panel (d) respectively. 
    The color scheme, line types, and labels have the same meaning as in Fig.\ref{fig:129124_spk_sequence} and Fig.\ref{fig:128125_spk_sequence}.}
    \label{fig:127126_spk_sequence}
\end{figure*}
\subsection{Irreducible $p/q$ tongues with multiple spike sequences}
\label{subsec:tongues_with_multiple_Spk}
Next, we describe the structure of $p/q$ tongues with multiple valid spike sequences. We elucidate the organization of spike sequences and bifurcations within the $2/13$ tongue, serving as a model for other irreducible tongues featuring numerous valid spike sequences. Out of six potential spike sequences $2/13$ frequency locking has  three valid spike sequences which are $\{1,2^4,1,2^9\},\{1,2^5,1,2^8\}$ and $\{1,2^6,1,2^7\}$. Fig.\ref{fig:2by13_external} shows the outer boundaries of each spike sequence in parameter space. The existence regions of three different spike sequences are seen to intersect. These regions have both pairwise and three way overlap. We computed the eigenvalues for each solution within these regions. We found that every spike sequence present in the region of overlap has atleast one stable solution, baring within small \lq U' regions. This leads to the emergence of \textit{intersequence} multistability. Segments of the outer boundaries of the three spike sequences join to yield the region of existence of $2/13$ frequency locking.
\par
In Fig.\ref{fig:2by13_external} each spike sequence possesses two kinds of outer and inner boundaries. The first kind is a bifurcation boundary. In our study, we find that these boundaries involve solutions of a single spike sequence. Solutions
belonging to different spike sequences do not interact. We term the second kind of boundary as a \lq spike exchange boundary'. This boundary is shared between two spike sequences. From Fig.\ref{fig:2by13_external} we observe that the spike sequences across this boundary differ by an exchange of two neighboring spikes, each originating from two different neurons. For instance, the sequence $\{1,2^{6},1,2^{7}\}$ transitions to $\{1,2^{5},1,2^{8}\}$ across one of the boundaries, by a spike exchange. The sequence $\{1,2^5,1,2^8\}$ then transitions to $\{1,2^4,1,2^9\}$ as we lower $\alpha$. No transition occurs directly from  $\{1,2^{6},1,2^{7}\}$ to $\{1,2^{4},1,2^{9}\}$. Transitions in every tongue follow a rule: non-neighboring spike exchanges are not allowed. We provide an explanation for this rule.
\par
Let us consider a periodic spike sequence $\{s(0),\cdots,s(m),s(m+1),s(m+2),\cdots,s(p+q-1)\}$ of a $p/q$ frequency-locked solution. Assume that in the spike sequence $s(m)=2, s(m+1)=2$ and $s(m+2)=1$. Next, focus our attention on $m^{th}$ to $(m+2)^{th}$ spikes and closely examine the times, $\tau_1$ and  $\tau_{2}$, for neuron $1$ and $2$ respectively to reach the threshold between the $m^{th}$ and $(m+1)^{th}$ spikes. Between these two spikes, both due to neuron $2$, $\tau_1-\tau_2>0$. On varying a parameter, suppose $\tau_1-\tau_2$ varies smoothly leading to
$\tau_1-\tau_2\rightarrow 0^+$, followed by $\tau_1=\tau_2$ and then $\tau_1-\tau_2\rightarrow 0^-$. At this point neuron $1$ reaches the threshold before neuron $2$, and fires the $(m+1)^{th}$ spike, while neuron $2$ is infinitesimally close to the threshold and will thus fires the $(m+2)^{th}$ spike. As a result, a neighboring exchange of spikes can occur. Next, we show that non-neighboring spike exchanges cannot occur. Within a spike sequence, let us consider two consecutive spike triplets separated by few other spikes. In the first triplet $s(m)=2, s(m+1)=2$ and $s(m+2)=2$  and in the second triplet $s(m')=2, s(m'+1)=1, s(m'+2)=2$. If first triplet transitions to $2,1,2$ and second one to $2,2,2$ on changing a parameter, a non-neighbouring spike exchange would have occurred. However, by the previous arguments, the triplet $2,1,2$ can transition to $2,2,1$ but not to $2,2,2$. Therefore, non-neighboring spike exchanges cannot take place.
\par
The bifurcation structure, computed from continuation algorithms, corresponding to the three \textit{individual} spike sequences of $2/13$ are shown in Figs.\ref{fig:129124_spk_sequence}-\ref{fig:127126_spk_sequence}. The outer boundaries of spike sequences are 
dominated by smooth saddle-node, Type-1 grazing bifurcation and spike exchange boundaries. However, small segments of these boundaries are also formed by Type-2 and Type-3 grazing bifurcations. In contrast, boundaries of the lone spike sequence of $1/q$ and $q/(q+1)$ described in earlier section are formed only by Type-1 grazing and saddle-node bifurcation. Bifurcations on these outer boundaries typically create either a stable spike sequence
through a grazing bifurcation or a stable-unstable pair through a saddle node bifurcation. We observed this behavior consistently in the case of $1/q$ and $q/(q+1)$. For each $2/13$ spike sequence, atypical segments on outer boundary generate a lone unstable solution via grazing bifurcation. This forms regions (\lq U') in Fig.\ref{fig:129124_spk_sequence}-\ref{fig:127126_spk_sequence}  where only one unstable and no stable spike sequence of $2/13$ exists.
 On varying $\alpha$ or $g$ across this region away from the outer boundary, stable solutions are created either through a Type-1 or saddle-node bifurcation (see Fig.\ref{fig:129124_spk_sequence} (a) and Fig.\ref{fig:127126_spk_sequence} (c)) or the lone unstable solution bifurcates into a stable one (see Fig.\ref{fig:128125_spk_sequence} (d) and Fig.\ref{fig:127126_spk_sequence} (d)) without the involvement of any other $2/13$ spike sequence. Note that in the \lq U' regions, stable solutions of other $p/q$ or quasiperiodic solutions could be present.\par
The inner boundaries corresponding to the three spike sequences render a rich sub-structure as seen in Figs.\ref{fig:129124_spk_sequence}-\ref{fig:127126_spk_sequence}. Several bifurcations and spike exchange shape these regions. Saddle-node, grazing bifurcations of Type-1, Type-2b and spike exchanges are found on inner boundaries of all the three spike sequences while Type-2a grazing bifurcation is absent in one of the spike sequences ($\{1,2^4,1,2^9\}$). In contrast, inner boundaries of irreducible $p/q$ tongues with single spike sequence (discussed earlier), do not possess saddle-node bifurcation and Type-2a grazing bifurcation while spike exchanges are naturally absent. Boundaries of $2/13$ spike sequences 
 defined by each bifurcation type and spike exchanges are depicted by different colors in Fig.\ref{fig:129124_spk_sequence}-\ref{fig:127126_spk_sequence}. Whether the grazing bifurcations and spike exchange boundaries involve a stable or unstable solution of a spike sequence is also indicated in the figures. Note that Type-2 bifurcations can only involve an unstable spike sequence
 (see Sec.\ref{sec:Bifurcations_And_multistability}). We computed the number of solutions in every region and determined their stability. The number of solutions (stable $+$ unstable) of a given spike sequence varies from one to four (shown in figures). Furthermore, multiple stable solutions of the same spike sequence, with different temporal structures, are found to coexist within a few regions. We refer to this coexistence as \textit{intra-sequence multistability}. Such multistable regions are labeled as \lq M' in Figs.\ref{fig:129124_spk_sequence}-\ref{fig:127126_spk_sequence}. Number of stable spike sequences can also be figured out from these figures by counting stable solutions created on moving across the boundaries into a region \lq M'. In doing so it is also evident that various sequences of bifurcations lead into \lq M'.
\par
In Sec.\ref{sec:simulation_results} we reported our observation of a Farey tree arrangement of spike sequences found from numerical simulations at large $\alpha$ values. Our analytical solutions agree with these observations. Fig.\ref{fig:2by13_external} shows that a unique spike sequence exists at large $\alpha$, while at intermediate and lower $\alpha$ values this is not so. Moreover, among all possible spike sequences at large $\alpha$, the valid one, $\{1,2^{6},1,2^{7}\}$ aligns with the Farey arrangement of sequences. 

\subsection{Reducible $np/nq$ islands}
\label{subsec:reducible_p/q}
\begin{figure}[t]
    \centering
    \includegraphics[scale=0.4]{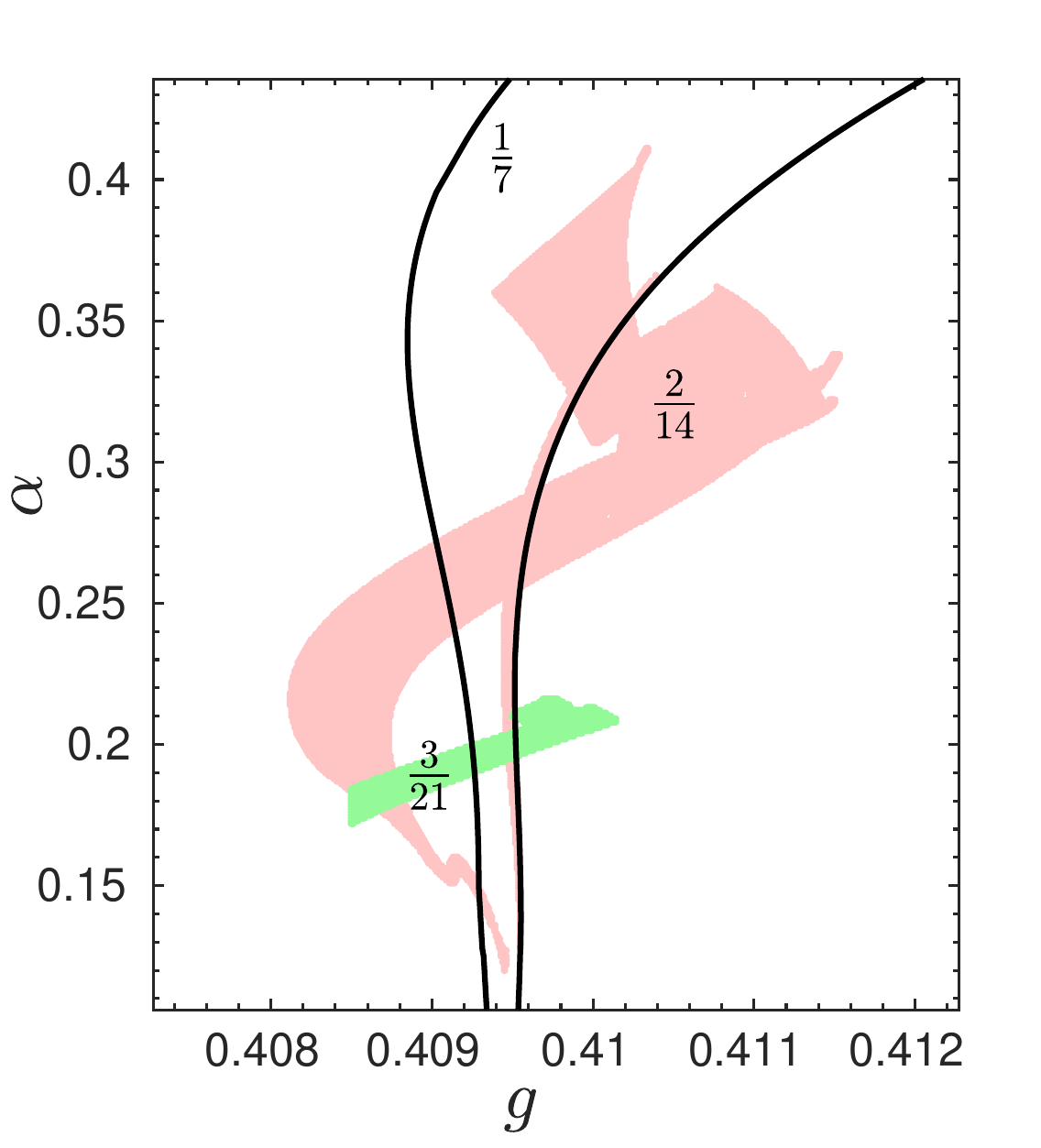}
    \caption{(Color online) Analytically obtained region of existence of $2/14$ island is shown. The numerically obtained $2/14$ island in Fig.\ref{fig:pspace_higher_resolution} ends abruptly whereas the analytically obtained $2/14$ island continues downwards. The $1/7$ tongue, $2/14$, and $3/21$ islands have a mutual overlap. The $1/7$ tongue is obtained analytically whereas $ 3/21$ is numerically obtained.}
    \label{fig:islands_mutual_overlap}
\end{figure}

In our numerical simulations, we observed reducible $np/nq$ frequency-locked regions manifesting as islands. We applied our analytical procedure to identify $2p/2q$ regions, revealing that these regions indeed appear abruptly as true islands. Fig.\ref{fig:islands_mutual_overlap} illustrates the frequency-locked island of $2/14$. $2/14$ possesses four valid spike sequences -$\{1,2^4,1,2^{10}\}$,$\{1,2^5,1,2^9\}$,$\{1,2^6,1,2^8\}$, and $\{1,2^{7},1,2^7\}$. The $2/14$ island, in Fig.\ref{fig:islands_mutual_overlap}, is formed from a union of regions bounded by the outer boundaries of all these spike sequences (not shown). Analysis for $2/10,2/12$, and $2/16$ cases also yielded similar island formation. We conjecture that all such $np/nq$ regions are organized as islands.
\par
Let us now focus on the ordering of frequency-locked regions of a family $np/nq$, $n\geq 1$. In Fig.\ref{fig:pspace_higher_resolution} of numerical simulations, we find that these regions appear to originate in increasing order of $n$ as $\alpha$ is lowered.  In the family $n/7n$, through a combination of analytical and numerical results, we find a mutual overlap of three successive regions $1/7,2/14$ and $3/21$ (shown in Fig.\ref{fig:islands_mutual_overlap}). We have observed a similar overlap in the family $n/6n$. From these results, along with similarities of the numerical results for various other $np/nq$ families to those of $n/6n$ and $n/7n$, it appears that the mutual overlap of family members of $np/nq$ is a general feature.
\par
We also conducted a bifurcation analysis for the $2p/2q$ islands specifically focusing on $2/12$ and $2/14$. The outer and inner boundaries exhibit similar bifurcation features as those observed in the case of irreducible tongues with multiple spike sequences. We did not carry out a similar analysis for $3p/3q$ islands and beyond, as the number of potential spike sequences increases significantly, making it difficult to fully probe the bifurcation structure of these islands.
\par
\begin{figure}[b]
    \centering
    \includegraphics[scale=0.45]{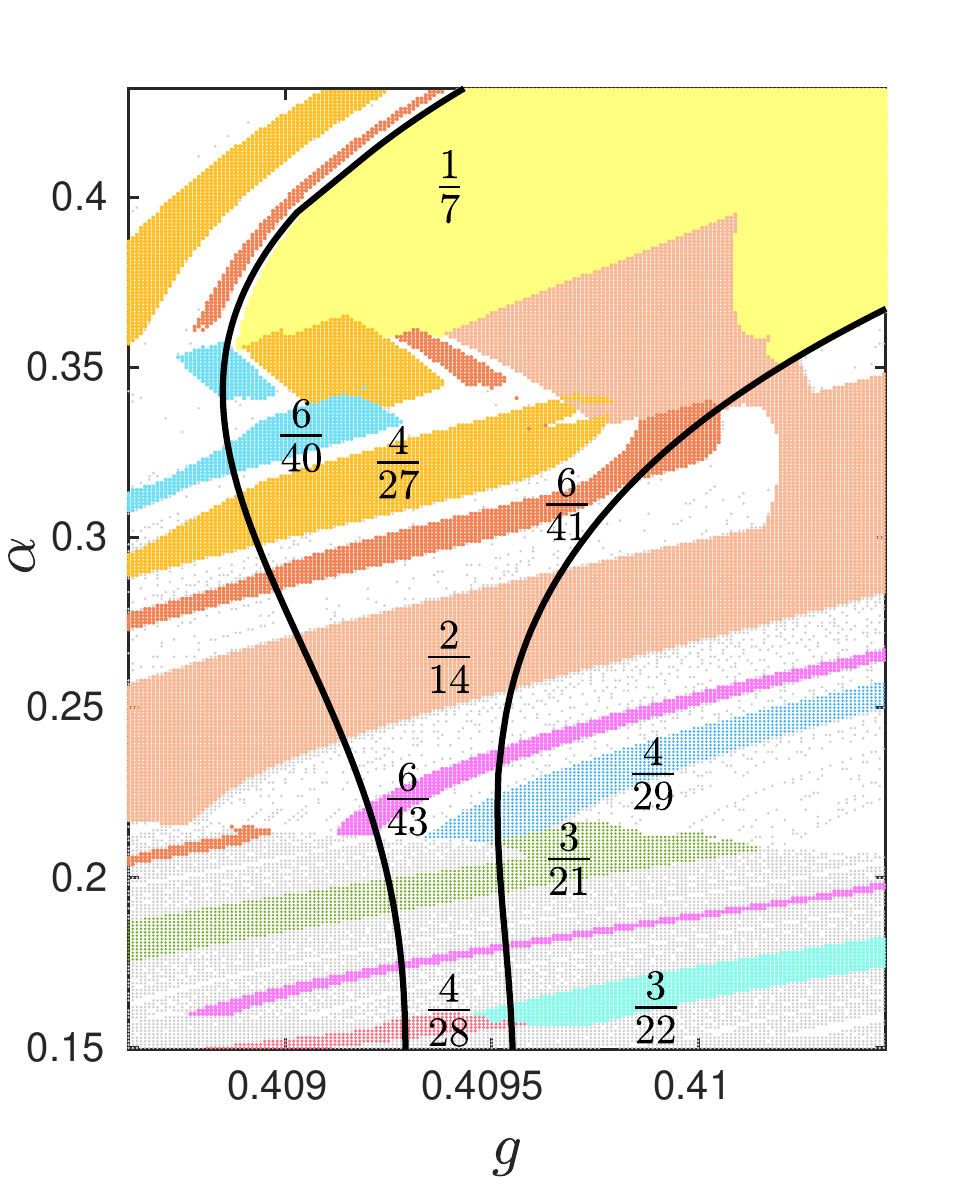}
    \caption{(Color online) Outer boundary of the $1/7$ tongue obtained from analytical method is shown in black. There exists one stable $1/7$ solution within the outer boundary. Grey represents quasiperiodic region. Other colors indicate various frequency-locked regions obtained from numerical simulations using the same initial condition. }
    \label{fig:1by7_Withothers}
\end{figure}
\subsection{Inter-mode multistability}
\label{subsec:pbq_And_others}
In Fig.\ref{fig:pspace_higher_resolution},  obtained through numerical simulations, we observed abruptly ending irreducible tongues and islands. Dominant examples of these include $p/q=1/6,1/7,1/8,2/13,2/15$ and several others. This abrupt ending arises due to the emergence of multistability, as demonstrated next through a combination of analytical and numerical results. In Fig.\ref{fig:1by7_Withothers} we present the analytically obtained boundary of the $1/7$ tongue along with the several numerically generated dominant tongues and islands. Contrary to simulation results, the $1/7$ tongue does not end abruptly but 
 continues downwards with decreasing width. Eigenvalue calculation shows that a stable solution exists throughout the tongue. Many numerically obtained regions overlap with the $1/7$ tongue, indicating the coexistence of $p/q=1/7$ with other $p'/q'$ attractors. Numerically found quasiperiodic regions also intersect with $1/7$ tongue and are thus multistable with it. We refer to multistability between two separate $p/q$ and $p'/q'$ attractors as \textit{inter-mode multistability}. This is the commonly found form of multistability among frequency-locked regions, as in the circle maps \cite{pikovsky2001synchronization}.
To further explore the generality of inter-mode multistability between $p/q$ and $p'/q'$, we analytically identified existence regions of several other irreducible $p/q$ tongues - $1/6,1/7,1/8,2/11,2/13$, and $2/15$. Each of these tongues starts from large $\alpha$ and continues downwards without interruption. A similar analysis for reducible $np/nq$ islands indicates 
 that their abrupt endings observed in numerical simulations are also due to multistability between the island with other frequency-locked regions. Therefore, the abrupt ending of frequency-locked regions or their reappearance as islands in simulations can now be understood as artefacts of initial conditions chosen in the multistable region. 
\section{Conclusion}
Our work shows that a pair of excitatory-inhibitory LIF neurons with \textit{identical} frequencies can exhibit a rich structure of high-order frequency locking, a phenomenon previously thought to require two \textit{dissimilar} frequencies. Instead, $1/1$ frequency locking present in identical neurons with symmetric coupling \cite{van_vreeswijk_1994} is absent 
under asymmetric coupling induced by the E-I pair. By systematically varying the parameters $g$ and $\alpha$ we found frequency-locked regions and quasiperiodicity. At large $\alpha$ values, the frequency-locked regions were characterized by devil’s staircase and Farey tree arrangement. Both these characteristics are well known in the case of mutually coupled non-identical oscillators and forced oscillators with frequency mismatches. In our work we found that Farey arrangement of $p/q$ tongues is accompanied by a Farey arrangement of spike sequences, not yet found in any previous study. This is intriguing because, even though multiple spike sequences are present in these $p/q$ frequency lockings, yet the one that adheres to the Farey arrangement is selected at large $\alpha$ values. At smaller $\alpha$ values the dynamics is more complex. Farey arrangement does not hold. Various $p/q$ regions overlap.  We found several sequences of reducible $np/nq$ islands created in increasing order of $n$ as $\alpha$ decreases.
\par
 The frequency-locked regions have an internal bifurcation structure, leading to the creation of new spike sequences with the same $p/q$ ratio or new solutions of the same spike sequence. In contrast, frequency-locked regions typically described in forced or coupled oscillators are structure-less \cite{pikovsky2001synchronization}. We found the bifurcation boundaries to be formed by saddle-node and non-smooth grazing bifurcations. The first of this is well known in forced oscillators \cite{pikovsky2001synchronization}.
Additionally, we found inter-mode ($p/q$ with $p'/q'$),  inter-sequence, and intra-sequence multistability. The first of these is well known in forced oscillators \cite{Coombes2001,KhajehAlijani2009}. However, the latter two can arise only through internal bifurcations within a $p/q$ region. Through our analysis we find that various $p/q$ regions (except $0/1$) share some common features. Outer boundaries of all $p/q$ regions contain saddle-node and Type-1 grazing bifurcations. Whereas, Type-1 and Type-2b bifurcations exists on inner boundaries of all the regions. Usually, a Type-2b grazing bifurcation boundary lies next to a saddle-node bifurcation, destroying the unstable spike sequence. The bifurcation structure for various $p/q$ regions, otherwise, differs in its details. The boundary of $0/1$ is defined entirely by Type-3 grazing bifurcation.
\par
Bifurcation structure was found by means of analytical method devised in the paper.
We employed event driven maps  to determine existence and linear stability for each spike sequence of a $p/q$ frequency locking. For determining existence, we constructed higher-order iterates of the event-driven maps of LIF neurons for a chosen spike sequence in terms of unknown interspike intervals. Imposing conditions of periodicity gave equations for the state vector of the two-neuron network in terms of interspike intervals. Solutions for state vectors and interspike intervals were thereafter found self-consistently. Additional conditions were imposed, considering the non-monotonicity of voltage flow of the neuron receiving inhibitory current, to remove unphysical solutions. For linear stability we constructed a matrix connecting perturbations separated by $(p+q)$ iterates. Our approach successfully removes unphysical solutions, is not constrained by coupling strength or the requirement of neurons being non-identical. Furthermore, it can identify non-smooth grazing bifurcations in coupled neurons, which were so far known to exist in forced neurons. In this context we also looked for grazing bifurcations in a pair of mutually pulse coupled identical inhibitory neurons studied in \cite{van_vreeswijk_1994}. These display $1/1$ frequency locking where an anti-sync state becomes stable through a pitchfork bifurcation. We discovered (not reported here) that the unstable branches disappears through a grazing bifurcation. In our work, we also found two new types of grazing bifurcations (Type-2b and Type-3), not reported earlier to best of our knowledge. 
\par
Our analytical framework can be further extended to explore periodic patterns of spike sequences in larger neuronal networks. The theoretical framework in \cite{Coombes1999} was bound to frequency locking in two-neuron networks. Excitatory-Inhibitory neuronal connections leads to a rich structure of spike patterns, as we found in our work. Such connections are known to be prevalent in the nervous system \cite{Reyes1998,Lin2024}. Specifically, in hippocampus and cortex, there are populations of E-I neurons with reciprocal interaction which generate periodic activity\cite{Traub1996,Buzski2002}. The principles characterizing and shaping the collective dynamics of single E-I pair of our work could provide valuable insights into oscillatory activity of different types in these regions of the brain.
Spike sequences are central to neural coding \cite{Ndasdy2000}, play a role in memory consolidation \cite{Carr2011}. 
Internal dynamics of neural circuits, characterized by excitatory-inhibitory connections and varied spike patterns, are relevant in this regard.
\par
This work extends the exploration of synchronization, suggesting that high-order frequency locking may occur in other coupled identical oscillators with some form of asymmetric coupling. In the context of neurons, question emerges: is the complex high-order frequency locking structure found in our work generic to other neuron models with excitatory-inhibitory coupling ? Our analytical framework could be extended to address this question in other integrate-and-fire models. For instance, an integrate-fire-or-burst model is known to reproduce the dynamics of relay neurons in the thalamus \cite{Knight1972}, ghostbursting neurons describing the dynamics of pyramidal cells in electro-sensory lateral line lobe of fish \cite{Doiron2002}. Though frequency locking under periodic forcing of a single neuron have been explored in these models \cite{Coombes2001,LAING2005} exploring their dynamics in a network is of relevance in neuroscience.

\vspace*{10pt}

\appendix

\section{Derivations of expressions for  $Q_i^+(n+1)$, $E^+_i(n+1)$ and $x^+_{i}(n+1)$ of Sec.\ref{sec:pbyq_freq_locked_analytical}}
\label{appendix:steps_For_calculations}
To obtain the expression for $Q^+_{i}(n+1)$, we put $n=0,1$ in Eq.(\ref{Eqn:Q_plus_map}) accompanied by Eq.(\ref{Eqn:Q_minus_map}) we get
\begin{align}
    Q^{+}_{i}(1)&=Q^{+}_{i}(0)e^{-\alpha\tau(0)}+\alpha^{2}C_{s(1),i} \quad \text{,}
    \label{Eqn:Qi(1)_plus_in_terms_of_Qi(0)}
\end{align}
\begin{subequations}
\begin{align}
    Q^{+}_{i}(2) &= Q^{+}_{i}(1)e^{-\alpha\tau(1)} + \alpha^{2}C_{s(2),i}
    \label{Eqn:Qi(2).1}
        \\
     \nonumber
    &= Q^{+}_{i}(0)e^{-\alpha(\tau(0)+\tau(1))} + \alpha^{2}C_{s(1),i}e^{-\alpha\tau(1)}\\ 
        &+\alpha^{2}C_{s(2),i}
    \label{Eqn:Qi(2).2}
\end{align}
\end{subequations}
Eq.(\ref{Eqn:Qi(2).2}) is obtained by substituting Eq.(\ref{Eqn:Qi(1)_plus_in_terms_of_Qi(0)}) into Eq.(\ref{Eqn:Qi(2).1}). Likewise \
\begin{align}
 Q^{+}_{i}(3) &= Q^{+}_{i}(0)e^{-\alpha(\tau(0) + \tau(1) + \tau(2))} \nonumber \\
 &\quad+ \alpha^{2}C_{s(1),i}e^{-\alpha(\tau(1) + \tau(2))} \nonumber \\
 &\quad + \alpha^{2}C_{s(2),i}e^{-\alpha\tau(2)} 
 + \alpha^{2}C_{s(3),i}
\end{align}

The above expression when generalized for any $n$ gives Eq.(\ref{Eqn:Qi(n+1)_plus_in_terms_of_Qi(0)_plus}). Expression for $E^+_{i}(n+1)$ can also be similarly obtained by repeated application of Eq.(\ref{Eqn:E_plus_map}) and (\ref{Eqn:Q_plus_map}). To obtain the expression for $x_{i}^+(n+1)$ we start with Eq.(\ref{Eqn:xi(1)plus})
\begin{align}
\nonumber
    x_{i}^{+}(1)&=C_{s(1),i}\Big[C_{s(0),i}x_{i}^{+}(0)e^{-\tau(0)}\\
    &+\zeta_{i}(\tau(0),E^{+}_{i}(0),Q^{+}_{i}(0))\Big]
    \label{Eqn:xi(1)plus_appendix}
\end{align}
Substituting $n=1$ in Eq.(\ref{Eqn:xi(n+1)plus}) we get
\vspace{-30pt}
\begin{widetext}
\begin{subequations}
\begin{align}
x_{i}^{+}(2) &= C_{s(2),i}\Big[x_{i}^{+}(1)e^{-\tau(1)} 
   + \zeta_{i}(\tau(1),E^{+}_{i}(1),Q^{+}_{i}(1))\Big]\label{Eqn:xi(2)plus_appendix}\\ 
   \nonumber
 & = x_{i}^{+}(0)e^{-(\tau(0)+\tau(1))}\prod_{l=0}^{2}C_{s(l),i}  + \zeta_{i}(\tau(0),E^{+}_{i}(0),Q^{+}_{i}(0))e^{-\tau(1)}\prod_{l=1}^{2}C_{s(l),i} \\
 &\qquad + \zeta_{i}(\tau(1),E^{+}_{i}(1),Q^{+}_{i}(1))\prod_{l=2}^{2}C_{s(l),i}
    \label{Eqn:xi(2)plus.2_appendix} 
\end{align}
\end{subequations}
Eq.(\ref{Eqn:xi(2)plus.2_appendix}) is obtained by substituting Eq.(\ref{Eqn:xi(1)plus_appendix}) in Eq.(\ref{Eqn:xi(2)plus_appendix}). Likewise 
\begin{align}
\nonumber
    x_{i}^{+}(3) &= C_{s(3),i}x_{i}^{+}(2)e^{-\tau(2)} + C_{s(3),i}\zeta_{i}(\tau(2),E^{+}_{i}(2),Q^{+}_{i}(2))\\
    \nonumber
    &= x_{i}^{+}(0)e^{-(\tau(0)+\tau(1)+\tau(2))}\prod_{l=0}^{3}C_{s(l),i} + \zeta_{i}(\tau(0),E^{+}_{i}(0),Q^{+}_{i}(0))e^{-(\tau(1)+\tau(2))}\prod_{l=1}^{3}C_{s(l),i} \\
    \nonumber
    &+ e^{-\tau(2)}\zeta_{i}(\tau(1),E^{+}_{i}(1),Q^{+}_{i}(1))\prod_{l=2}^{3}C_{s(l),i}+ C_{s(3),i}\zeta_{i}(\tau(2),E^{+}_{i}(2),Q^{+}_{i}(2))\\
    \nonumber
    &= x_{i}^{+}(0)e^{-\Delta(0,2)}P_{i}(0,3)+ \zeta_{i}(\tau(0),E^{+}_{i}(0),Q^{+}_{i}(0))e^{-\Delta(1,2)}P_{i}(1,3) \\
    & + \zeta_{i}(\tau(1),E^{+}_{i}(1),Q^{+}_{i}(1))e^{-\Delta(2,2)}P_{i}(2,3) + P_{i}(3,3)\zeta_{i}(\tau(2),E^{+}_{i}(2),Q^{+}_{i}(2))
\end{align}
\end{widetext}
 Generalizing above expression for any $n$ we get Eq.(\ref{Eqn:xi(n+1)_plus_in_terms_of_xi(0)_plus}).
\section{Firing death boundary - analytical solution}
\label{appendix:Firing_Death}
In this appendix, we describe the application of the procedure described in  Sec.\ref{sec:pbyq_freq_locked_analytical} to the case of firing death in an E-I pair. We begin from step $2$ of the procedure.\\ \\
\textbf{Step 2}: In the firing death state the spike sequence $\{s(0),s(1),\cdots\}=\{2,2,\cdots\}$. While neuron $1$ does not fire any spikes, neuron $2$ evolves as a free LIF neuron with its voltage rising monotonically towards the threshold. The state of the system is the same at every spike which from Eqs.(\ref{Eqn:Qi(0)tilda})-(\ref{Eqn:xi(0)tilda}) is
\begin{equation}
    \tilde{Q}^{+}_{i}(0)=\frac{\alpha^2C_{2,i}}{\Big(1-e^{-\alpha\tilde\tau(0)}\Big)} \quad,
    \label{Eqn:FD_Qi(0)}
\end{equation}
\begin{equation}
    \tilde{E}^{+}_{i}(0)=\frac{\tilde{Q}^{+}_{i}(0)\tilde{\tau}(0) e^{-\alpha\tilde{\tau}(0)}}{\Big(1-e^{-\alpha\tilde{\tau}(0)}\Big)} \quad,
\label{Eqn:FD_Ei(0)}
\end{equation}
and
\begin{equation}
    \tilde{x}^{+}_{i}(0)=\frac{P_{i}(1,1)\zeta_{i}(\tilde{\tau}(0),\tilde{E}^{+}_{i}(0),\tilde{Q}^{+}_{i}(0))}{\Big(1-e^{-\tilde{\tau}(0)}P_{i}(0,1)\Big)} \quad,
    \label{Eqn:FD_xi(0)}
\end{equation}
where $P_{i}(1,1)=\prod\limits_{l=1}^{1}C_{s(l),i}=C_{2,i}$ and $P_{i}(0,1)=\prod\limits_{l=0}^{1}C_{s(l),i}=C_{2,i}C_{2,i}$. Since $C_{2,1}=1$ and $C_{2,2}=0$ it follows that 
$P_{1}(1,1)=1$, $P_{2}(1,1)=0$, $P_{1}(0,1)=1$, and $P_{2}(0,1)=0$ which yields for neuron $1$
\begin{equation}
 \tilde{Q}^{+}_{1}(0)= \frac{\alpha^{2}}{\Big(1-e^{-\alpha \tilde{\tau}(0)}\Big)}  \quad,
 \label{Eqn:Q1(0)_firingdeath}
\end{equation}
\begin{equation}
    \tilde{E}^{+}_{1}(0)=\frac{\tilde{Q}^{+}_{1}(0)\tilde{\tau}(0) e^{-\alpha \tilde{\tau}(0)}}{\Big(1-e^{-\alpha\tilde{\tau}(0)}\Big)} \quad,
    \label{Eqn:E1(0)_firingdeath}
\end{equation}
and
\begin{equation}
\tilde{x}^{+}_{1}(0)= \frac{\zeta_{1}(\tilde{\tau}(0),\tilde{E}^{+}_{1}(0),\tilde{Q}^{+}_{1}(0))}{\Big(1-e^{-\tilde{\tau}(0)}\Big)} \quad,
\label{Eqn:x1(0)_firingdeath}
\end{equation}
For neuron $2$
\begin{equation}
    \tilde{Q}^{+}_{2}(0)=0 \quad,
\label{Eqn:FD_Q2(0)plus}
\end{equation}
\begin{equation}
    \tilde{E}^{+}_{2}(0)=0 \quad,
    \label{Eqn:FD_E2(0)plus}
\end{equation}
and
\begin{equation}
    \tilde{x}^{+}_{2}(0)=0 \quad.
    \label{Eqn:FD_x2(0)plus}
\end{equation}
The values of the synaptic current variables $\tilde{Q}_{2}^{+}(0)$ and $\tilde{E}_{2}^{+}(0)$  are both zero, as expected, since neuron $1$ having ceased to fire does not send any current into neuron $2$. Likewise, $x_{2}^{+}(0)$ was also expected to be equal to zero, since neuron $2$ fires all spikes. \\ \\
\textbf{Step 3}:  $\tilde{\tau}(0)$ should satisfy Eq.(\ref{Eqn:tau_eqns})
\begin{equation}
    \tilde{x}^{+}_{2}(0)e^{-\tilde{\tau}(0)}+\zeta_{2}(\tilde{\tau}(0),\tilde{E}^{+}_{2}(0),\tilde{Q}^{+}_{2}(0))=1
    \label{Eqn:tau_for_Firing_neuron_FD} ,
\end{equation}
where  
\begin{align*}
    \zeta_{2}(\tilde{\tau}(0),\tilde{E}^{+}_{2}(0),\tilde{Q}^{+}_{2}(0))=a(1-e^{-\tilde{\tau}(0)})\\+g_{2}\Bigg[\frac{e^{-\tilde{\tau}(0)}-e^{-\alpha\tilde{\tau}(0)}}{\alpha-1}
    \Big(\tilde{E}^{+}_{2}(0)+\frac{\tilde{Q}^{+}_{2}(0)}{\alpha-1}\Big)\\
  -\frac{\tilde{Q}^{+}_{2}(0)\tilde{\tau}(0)e^{-\alpha\tilde{\tau}(0)}}{(\alpha-1)}\Bigg]
\end{align*}
\textbf{Step 4:}
Substituting Eqs.(\ref{Eqn:FD_Q2(0)plus})-(\ref{Eqn:FD_x2(0)plus}) into Eq.(\ref{Eqn:tau_for_Firing_neuron_FD}) we get the functional form for $\mathcal{H}_{0}(\tilde{\tau}(0))$ as
\begin{equation}
    a(1-e^{-\tilde{\tau}(0)})-1=0 ,
\end{equation}
the solution of which gives 
\begin{equation}
    \tilde{\tau}(0)=\ln\Big(\frac{a}{a-1}\Big)
    \label{Eqn:tau0_firingdeath} ,
\end{equation}
and exists for all $(g,\alpha)$. The solution $\tilde{\tau}(0)$ is \textit{unique} and equal to the free time period of the neuron, as expected. Substituting $\tilde{\tau}(0)$ in Eqs.(\ref{Eqn:Q1(0)_firingdeath})-(\ref{Eqn:FD_x2(0)plus}) and  gives the fixed state of neuron $1$ at every spike of neuron $2$ while the state variables of neuron $2$ are anyway equal to zero.\par
The solutions, Eqs.(\ref{Eqn:Q1(0)_firingdeath})-(\ref{Eqn:FD_x2(0)plus}) and Eq.(\ref{Eqn:tau0_firingdeath}), obtained above exist for all $(g,\alpha)$ but will exist only if they satisfy conditions $1$ and $2$ of  Sec.\ref{sec:pbyq_freq_locked_analytical}. Let us examine these. \\
\begin{enumerate}
    \item Condition $1$: Starting from $\tilde{x}^{+}_2(0)=0$, the underlying voltage trajectory of neuron $2$ rises monotonically, crossing the threshold only once. Hence, $\tilde{\tau}(0)=\tau_2^*(0)$
  irrespective of $(g,\alpha)$, satisfying condition $1$.
    \item Condition $2$: $\tau_{1}^*(0)$ should be obtained from Eq.(\ref{Eqn:eqnfortau_i}) which takes the form
\begin{flalign}
       \hspace{0.6cm} \tilde{x}^{+}_{1}(0)e^{-\tau^*_{1}(0)}+\zeta_{1}(\tau^*_{1}(0),\tilde{E}^{+}_{1}(0),\tilde{Q}^{+}_{1}(0)) =1
\end{flalign}
Substituting $\tilde{\tau}(0)$ into Eqs.(\ref{Eqn:Q1(0)_firingdeath})-(\ref{Eqn:x1(0)_firingdeath})
and then placing those into the above equation yields a transcendental equation in $\tau^*_{1}(0)$.
This equation needs to be solved numerically for $\tau^*_{1}(0)$ for a given $(g,\alpha)$. For the firing death state to exist, $\tau^*_{1}(0)$ thus obtained should satisfy $\tau_{1}^*(0)>\tilde{\tau}(0)$. 
\end{enumerate}
Since the above solutions satisfy condition $1$ for all $(g,\alpha)$, the region of existence of firing death state is determined solely by condition $2$. Therefore, the state will get lost when condition $2$ is violated.

\par
Next, we determine the stability of the firing death state. We do this by calculating the eigenvalues of the Jacobian $\Big(\mathbf{DF}(0)\Big)_{2}$ of  Sec.\ref{sec:pbyq_freq_locked_analytical}. Since the Jacobian has an upper triangular form its eigenvalues are given by the diagonal elements of matrices $\mathbf{R_2}$ and $\mathbf{U_2}$. Finding the partial derivatives in the matrices $\mathbf{R_2}$ and $\mathbf{U_2}$ we get
\begin{equation}
    \mathbf{R_2}= \begin{bmatrix}
    0 & 0 & 0 \\
    0 & e^{-\alpha\tilde{\tau}(0)} & e^{-\alpha\tilde{\tau}(0)} \\
    0 & 0 & e^{-\alpha\tilde{\tau}(0)} 
    \end{bmatrix}  
\end{equation}
and
\begin{equation}
\mathbf{U_2}= \begin{bmatrix}
    e^{-\tilde{\tau}(0)} & -g\kappa_1(\tilde{\tau}(0)) & -g\kappa_2(\tilde{\tau}(0)) \\
    0 & e^{-\alpha\tilde{\tau}(0)} & e^{-\alpha\tilde{\tau}(0)} \\
    0 &0& e^{-\alpha\tilde{\tau}(0)} 
    \end{bmatrix},
\end{equation}
where $\kappa_1(\tau)=\frac{1}{\alpha-1}\big(e^{-{\tau}}-e^{-\alpha\tau}\big)$ and $\kappa_2(\tau)=\frac{1}{(\alpha-1)^2}\big(e^{-\tau}-e^{-\alpha\tau}\big)-\frac{1}{(\alpha-1)}\tau e^{-\alpha\tau}$.
Therefore eigenvalues of Jacobian $\Big(\mathbf{DF}(0)\Big)_2$ are  $e^{-\alpha\tilde{\tau}(0)}$ with multiplicity of four, $e^{-\tilde{\tau}(0)}$ and $0$. Since $\tilde{\tau}(0)=\ln\Big(\frac{a}{a-1}\Big)$ magnitude of all the eigenvalues of the Jacobian are less than one. Hence the firing death state is stable in the region of parameter space where it exists.\\\\
We will proceed with the knowledge that violation of condition $2$ occurs through Type-$3$ grazing bifurcation (see Sec.\ref{sec:pbyq_freq_locked_analytical}). At the Type-3 bifurcation point, in the case of firing death, neuron $1$ should be at the threshold, implying 
\begin{align}
  x_1^-(\tau^*_1(0))=  \mathcal{F}_1(\tilde{x}_1^{+}(0),\tilde{E}_1^{+}(0),\tilde{Q}_1^{+}(0),\tau_1^*(0))=1,
    \label{Eqn:FD_eqn_for_tau1}
\end{align}
and $\frac{dx^-_1}{du}\Big\rvert_{\tau^*_1(0)}=0$ leading to
\begin{align}
  a-1-gE_1(\tau_1^*(0))=0
    \label{Eqn:dx1dt_at_tau1star}
\end{align}
where $E_{1}(\tau^*_1(0))=\big(\tilde{E}^{+}_{1}(0)+\tilde{Q}^+_{1}(0)\tau_1^*(0)\big)e^{-\alpha\tau_1^*(0)}$ from Eq.(\ref{Eqn:E_minus_map}). Additionally, Type-3 bifurcation requirement $\frac{dx_2^-}{du} \Bigg\rvert_{\tilde{\tau}(m)}\neq 0$
is trivially ensured, since voltage of neuron $2$ increases monotonically.

We make a few observations that will guide us in introducing a suitable approximation. Differentiating Eq.(\ref{Eqn:LIFmodeleqn}) in terms of $u$ for neuron $1$ we get 
\begin{align}
    \frac{d^2x_1}{du^2}=-\frac{dx_1}{du}- g\frac{dE_1}{du}\quad.
\end{align}
It follows that, at the bifurcation point,  $\frac{dE_1}{du}\Big\rvert_{\tau_1^*(0)}>0$. Hence the graze can occur only during the phase of increasing 
synaptic current. This implies that $\tau_1^*(0)$, at the bifurcation point, will occur before the maxima of synaptic current which can be shown to be located at $\tau_{max}=\frac{1}{\alpha}-\frac{E_1^{+}(0)}{Q_1^{+}{0}}$. Since $\tau_{max}>0$ it implies that $\frac{E_1(0)}{Q_1{0}}\alpha<1$. Hence,  $\tau_1^*(0)<\frac{1}{\alpha}-\frac{E_1(0)}{Q_1{0}}$ yielding $\alpha\tau_1^*(0)<1$. Let us approximate 
$e^{-\alpha\tau_1^*(0)}\approx1-\alpha\tau_1^*(0)+\frac{{(\alpha\tau_1^*(0))}^2}{2}$. Substituting this in Eq.(\ref{Eqn:dx1dt_at_tau1star}) and retaining terms up to second order in $\tau_{1}^*(0)$ we get the following quadratic equation
\begin{align}
\nonumber
   \Big[g\alpha\tilde{Q}_1^{+}(0)-\frac{ g\tilde{E}_1^{+}(0)\alpha^2}{2}\Big] {\tau_{1}^*(0)}^2 \\ +\tau_1^*(0)\Big[g\alpha\tilde{E}_1^+(0)-g\tilde{Q}_1^{+}(0)\Big]+a-1-g\tilde{E}^+_1(0)&=0.
\label{Eqn:quadaratic_eqn_in_tau1star}
\end{align}
Its solution, $\tau_1^*(0)$, using Eq.(\ref{Eqn:Q1(0)_firingdeath}) and (\ref{Eqn:E1(0)_firingdeath}), can be expressed entirely as a function of $g$ and $\alpha$. Substituting the resulting solution for $\tau_1^*(0)$ and  $\tilde{Q}_{1}^{+}(0)$, $\tilde{E}_1^+{0}$, $\tilde{x}^+_1(0)$ from Eqs.(\ref{Eqn:Q1(0)_firingdeath})-(\ref{Eqn:x1(0)_firingdeath})  into Eq.(\ref{Eqn:FD_eqn_for_tau1})  we get an implicit equation of the form $\uppsi(g,\alpha)=0$. This equation describes the Type-3 grazing bifurcation boundary of firing death solution in $(g,\alpha)$ parameter space.

\end{document}